\documentclass[12pt]{article}
\usepackage{setspace}
\usepackage{comment}
\usepackage{natbib}
\usepackage{graphicx}
\usepackage{lscape}
\usepackage{pdflscape}
\usepackage{afterpage}
\usepackage{pdfpages}
\usepackage{float}
\usepackage{booktabs}
\usepackage[utf8]{inputenc}
\usepackage{indentfirst}
\usepackage{arydshln}
\usepackage{arydshln}
\usepackage{tikz}
\usepackage{lipsum}
\usepackage{pgffor}
\usepackage{dsfont}
\usepackage{amsfonts}
\usepackage{amsmath}
\usepackage{xfrac}
\usepackage{amssymb}
\usepackage{graphicx}
\usepackage{url}				
\usepackage{subfig} 

\usepackage{array}            
\newcolumntype{C}[1]{>{\centering\arraybackslash}p{#1}}

\usepackage{threeparttable}   
\usepackage{float}            
\usepackage{amsmath}          

\usepackage[linesnumbered,ruled,vlined]{algorithm2e}
\usepackage{graphicx}

\newtheorem{proposition}{Proposition}[section]
\newtheorem{remark}{Remark}

\newenvironment{proof}[1][Proof]{\noindent \textbf{#1.} }{\ \rule{0.5em}{0.5em}}

\usepackage{hyperref} 
\hypersetup{
	colorlinks=true, breaklinks=true, bookmarks=true,bookmarksnumbered,
	urlcolor=red, linkcolor=red, citecolor=blue, 
	pdftitle={}, 
	pdfauthor={\textcopyright}, 
	pdfsubject={}, 
	pdfkeywords={}, 
	pdfcreator={pdfLaTeX}, 
	pdfproducer={LaTeX with hyperref and ClassicThesis} 
}

\usepackage[top=1in, bottom=1in, left=1in, right=1in]{geometry}

\usepackage{bbm}

\begin{document}

	\def\spacingset#1{\renewcommand{\baselinestretch}%
		{#1}\small\normalsize} \spacingset{1}

	
\title{ {
\large ASSESSING INFERENCE METHODS}\footnote{I would like to thank Alberto Abadie, Arun Advani, Isaiah Andrews, Luis Alvarez, Kirill Borusyak, Carol Caetano, Greg Caetano, Brant Callaway, Aureo de Paula, Christian Hansen, Xavier D'Haultfoeuille, Marcelo Fernandes, Avi Feller, Lucas Finamor, Peter Hull, Toru Kitagawa, Michael Leung, Marcelo Medeiros, Duda Mendes, Marcelo Moreira, Vitor Possebom, Marcelo Sant'Anna, Pedro Sant'Anna, Andres Santos, Jesse Shapiro, Rodrigo Soares, and Chris Walters for excellent comments and suggestions. I also thank participants at the SBE Applied Micro Seminar, PUC-Chile, Chicago, PIMES, University of Georgia, and Erasmus University seminars, and at the EEA, RES and SBE conferences. Luis Alvarez, Lucas Barros, Raoni Oliveira and Davi Siqueira provided exceptional research assistance. I also thank Pedro Ogeda for discussing with me an application that led me to think about this assessment for the first time. This paper previously circulated with the title ``A simple way to assess inference methods.'' I gratefully acknowledge financial support from FAPESP and CNPq. }}

\author{
Bruno Ferman\footnote{email: bruno.ferman@fgv.br; address: Sao Paulo School of Economics, FGV, Rua Itapeva no. 474, Sao Paulo - Brazil, 01332-000; telephone number: +55 11 3799-3350} \\
\\
Sao Paulo School of Economics - FGV \\
\footnotesize
First Draft: December 15th, 2019 \\
\footnotesize
This Draft: October 2nd, 2025
}

\date{}
\maketitle

	\newsavebox{\tablebox} \newlength{\tableboxwidth}
	

	\begin{center}

\textbf{Abstract}

\end{center}

We analyze different types of simulations that applied researchers can use to assess whether their inference methods reliably control false-positive rates. We show that different assessments involve trade-offs, varying in the types of problems they may detect, finite-sample performance, susceptibility to sequential-testing distortions, susceptibility to cherry-picking, and implementation complexity. We also show that a commonly used simulation to assess inference methods in shift-share designs can lead to misleading conclusions and propose alternatives. Overall, we provide novel insights and recommendations for applied researchers on how to choose, implement, and interpret inference assessments in their empirical applications.

\

	\noindent%
	
	\
	
	\noindent%
		
		\vfill

	\newpage
	\spacingset{1.45} 
	

\doublespace

\section{Introduction}

The credibility of scientific research depends crucially on the control of false-positive results. However, we may have an excess of false-positive results if inference is based on incorrect/unreliable methods. This may happen when (i) inference methods rely on unrealistic assumptions, (ii) inference is based on asymptotic theory when such asymptotic approximations are poor, and/or (iii) the inference method is invalid even asymptotically.

We analyze different types of simulations that applied researchers may use to assess whether inference methods in their empirical applications reliably control for false-positive rates. The main idea is to consider a data generating process (DGP) tailored to the empirical application at hand in which the null hypothesis is true, and then compute the rejection rate of a given inference method under this DGP. While there have been a number of papers proposing the use of simulations tailored to the data at hand to evaluate statistical methods,\footnote{A non-exhaustive list of papers that consider the use of simulations tailored to the data at hand includes \cite{HuberLechnerWunsch2013,Huber_EMC}, \cite{Busso}, \cite{athey2020using}, \cite{blair_cooper_coppock_humphreys_2019}, \cite{gelman2020bayesian}, and \cite{Morris}. \cite{Advani} is most related to our paper, as they analyze the use of different types of simulations applied researchers may use to select among treatment effect estimators under unconfoundedness. We discuss in Section \ref{Sec: DID comparison with Advani} how our conclusions differ from theirs.} we show that the use of simulations by applied researchers as inference assessments in empirical applications involves subtle issues that have not been previously considered in the literature. These issues revolve around the trade-offs between different approaches and the validity of their implementation. 

We consider  in Section \ref{Sec: DiD} the use of ``model-based'' assessments, in which the DGP used in the simulations holds covariates fixed and treats errors as stochastic. In the context of Difference-in-Differences (DiD), we analyze the use of different model-based assessments, which differ in how the error distribution in the simulations is specified. These alternatives range from simple DGPs with iid normal errors to more complex ones where the error distribution is estimated from the data. We show that these alternative assessments present non-trivial trade-offs, including the types of problems they may detect, finite-sample performance, susceptibility to sequential-testing distortions, cherry-picking opportunities, and implementation complexity. 

In Section \ref{Sec: shift-share}, we turn to ``design-based'' assessments, which use a DGP that holds potential outcomes fixed while treating covariates as stochastic, with a focus on shift-share design applications. This type of simulation was used by \cite{Adao} to show that inference based on robust or cluster-robust standard errors can lead to substantial over-rejection when errors are spatially correlated in shift-share designs. We formally demonstrate that this procedure may over-state the spatial correlation problem when a true treatment effect is present in the empirical application, and we discuss alternatives that address this issue. We also show that design-based simulations can provide valuable information about the error distribution in the empirical application, even though they rely on DGPs in which errors are non-stochastic.

In Section \ref{Sec: implementation}, we discuss key implementation decisions that applied researchers face when designing inference assessments, such as which elements to treat as stochastic in the simulations and which significance levels to evaluate. We show that focusing on rejection rates at a single significance level can be misleading when assessing inference methods that rely on standard errors computed with null-imposed residuals. In such cases, the inference method may under-reject at some significance levels and over-reject at others. This insight is novel and has implications not only for inference assessments but also for Monte Carlo (MC) simulations more broadly.

Overall, we provide new insights on the use of simulations as inference assessments and provide empirical evidence on their value in applied work. Section \ref{Recommendations} concludes with practical recommendations on how researchers can choose, implement, and interpret inference assessments in their empirical applications.

\section{Difference-in-Differences: Model-Based Assessments}
\label{Sec: DiD}

\subsection{Setting}
\label{Sec: DID setting}

Consider an applied researcher running a DiD regression:
\begin{eqnarray} \label{Eq_DID}
Y_{it} = \beta D_{it} +  \theta_{i} + \gamma_t + \epsilon_{it}, 
\end{eqnarray}
where $Y_{it}$ represents the outcome for state $i \in \{1,...,N\}$ at time $t \in \{1,...,T\}$; $ \theta_{i}$  and $ \gamma_t $ represent, respectively, the state and time fixed effects, while  $\epsilon_{it}$ is the regression error. The dummy variable $D_{it}$ represents a state-wide policy, and we define $\beta$ as the average treatment effect on the treated (ATT). For simplicity, we assume that all treated states start treatment after period $t^\ast$. Let $\mathbf{Y}$ be the $N \times T$ vector stacking all information on $Y_{it}$ for all states and all periods.

In this section, we consider a setting in which treatment assignment is fixed, where states $i = 1,...,N_1$ are treated, while $i=N_1 + 1,...,N$  are controls,  so $N_0 = N-N_1$. In this case, uncertainty comes from unobserved shocks (for example, economic or weather shocks) as well as from treatment effect heterogeneity, comprising the error term $\epsilon_{it}$.  See \cite{AFW_Few_Treated} for further discussion on considering DiD applications in a model-based setting with fixed treatment assignment. The goal is to estimate and draw inference on $\beta$.

The standard DiD estimator is given by 
\begin{eqnarray}
\hat \beta_{\mbox{\tiny DiD}}  &=&  \frac{1}{N_1} \sum_{i=1}^{N_1} \left[  \frac{1}{T-t^\ast} \sum_{t=t^\ast+1}^T Y_{it} - \frac{1}{t^\ast} \sum_{t=1}^{t^\ast} Y_{it}   \right]   -    \frac{1}{N_0} \sum_{i=N_1+1}^{N} \left[ \frac{1}{T-t^\ast} \sum_{t=t^\ast+1}^T Y_{it} - \frac{1}{t^\ast} \sum_{t=1}^{t^\ast} Y_{it}   \right] \nonumber \\
& = &  \beta + \frac{1}{N_1} \sum_{i=1}^{N_1} W_i - \frac{1}{N_0} \sum_{i=N_1+1}^{N} W_i,
\end{eqnarray}
where $W_i = \frac{1}{T-t^\ast} \sum_{t=t^\ast+1}^T \epsilon_{it} - \frac{1}{t^\ast} \sum_{t=1}^{t^\ast} \epsilon_{it}$ is the post-pre difference in average errors in state $i$. We consider the case in which the parallel trends identification assumption for the DiD estimator is valid, so $\mathbb{E}[W_i] = 0$ for all $i=1,...,N$, implying that  $\mathbb{E}[\hat \beta_{\mbox{\tiny DiD}} ] = \beta$. 

Suppose we want to test a null hypothesis $H_0: \beta = 0$ versus $H_1 : \beta \neq 0$. Therefore, we consider an inference method that will determine, for each realization of the data $\mathbf{Y}$, whether to reject or not the null hypothesis.\footnote{For simplicity, we consider non-randomized tests, so that the probability of rejection for a given realization of $\mathbf{Y}$ is either 0 or 1, instead of being in the interval $[0,1]$. Also, given that we are considering a setting in which treatment assignment is fixed, the test decision is a function of only  $\mathbf{Y}$, instead of being a function of  outcomes and covariates.} One of the main goals of an inference method is to control for the probability of detecting false-positive results. That is, we want the probability of rejecting the null when the null is true not to be larger than a  pre-specified significance level $\alpha$. 

Let $\boldsymbol{\epsilon}$ be an $N \times T$ vector stacking information of $\epsilon_{it}$ on all states for all periods, and $F(.)$ be a cumulative distribution function (CDF) for this vector of random variables. For a specific test procedure, we define $\Gamma (F(.))$ as the  rejection probability under the null $\beta=0$, when $F(.)$ is the distribution of the errors.\footnote{In most cases, this probability will not depend on the values of the fixed effects. Therefore, we define $\Gamma(.)$ as a function of only $F(.)$ for ease of exposition. We clarify the conditions in which this is the case in Remark \ref{Remark: invariance}. } Also, for a family $\mathcal{F}$ of possible distributions $F(.)$,  the finite-sample size of the test relative to this class of distributions is given by $\sup_{F(.) \in \mathcal{F}} \Gamma(F(.))$. This gives the worst-case scenario for the probability of detecting a false-positive given this admissible set of distribution for the errors, $\mathcal{F}$. 
We generally have to impose restrictions on the family $\mathcal{F}$, otherwise it would usually be impossible to construct non-trivial tests that control for finite-sample size \citep{bahadur1956,Romano2004,BertanhaMoreira2020}.

The goal then is to choose an inference method such that $\sup_{F(.) \in \mathcal{F}} \Gamma(F(.))$ is close to $\alpha$ for a family of distributions $\mathcal{F}$ appropriate to the empirical application. 
Since \cite{Bertrand04howmuch}, it is common  to consider inference based on cluster-robust standard errors at the state level. This allows  for unrestricted correlation in $\epsilon_{it}$ within state $i$. It also allows for unrestricted heteroskedasticity, so the distribution of $W_i$ might be different for treated and control states. On the other hand, it relies on independence of errors  $\epsilon_{it}$ across states  and on an asymptotic approximation in which both $N_0$ and $N_1$ are large. Before \cite{Bertrand04howmuch}, it was common to rely on robust standard errors, which would rely on  asymptotic approximations in which  the number of  treated and control state $\times$ time cells are large. However, a major disadvantage is that this alternative would not allow for serial correlation. There are also other  alternatives that are valid in small $N_1$ settings, but they would generally require additional assumptions (see \cite{AFW_Few_Treated} for a survey).

Crucially, there are a number of different ways to conduct inference in this setting, where alternatives vary in the set of assumptions and on the asymptotic approximations they rely on. Therefore, applied researchers should be cautious to choose an inference method that relies on assumptions and asymptotic approximations that are reasonable  in their empirical applications, so that inference does not result in an excess of false-positive results. However, this is not always the case. 

First, applied researchers may use inference methods that rely on unreasonable assumptions in their settings. For example, \cite{Bertrand04howmuch} show that many papers used to rely on robust standard errors in DiD applications. The problem in this case is that this inference method implicitly assumes a family $\mathcal{F}$ that does not take into account the possibility of serial correlation, which is usually a relevant feature  in this setting. Second, asymptotic theories may provide poor approximations, as documented by \cite{Young} in experimental papers. In this case, even if we define a reasonable family $\mathcal{F}$ such that $\lim\sup_{N \rightarrow \infty} \Gamma(F(.)) \leq \alpha$, it may be that $N$ is not large enough, so  $\sup_{F(.) \in \mathcal{F}} \Gamma(F(.))$  may not be close to $\alpha$. Finally, applied researchers may consider inference procedures that are invalid even asymptotically (we present  examples in Section \ref{Sec: problems even with large N}). In this case, we would have that $\Gamma(F(.)) \not \rightarrow \alpha$ for any $F(.) \in \mathcal{F}$.

\subsection{Model-Based Assessments }

Consider  applied researchers who are running the DiD regression from Equation \ref{Eq_DID} and relying on a specific inference method (for example, a t-test using cluster-robust standard errors). We discuss alternative inference assessments they may use to evaluate whether such specific inference method is reliable for their empirical application.

In order to provide an assessment for this inference method, we consider its rejection rate under the null for a given  distribution for the errors  $\widetilde F(.;  \mathbf{Y})$, which might depend on the observed data $\mathbf{Y}$. That is,  $\Gamma(\widetilde F(.;  \mathbf{Y}))$. While, $\Gamma(\widetilde F(.;  \mathbf{Y}))$ can potentially be calculated analytically, it is usually more convenient  to rely on simulations. More specifically, we generate data from the DGP considered in Equation \ref{Eq_DID}, imposing the null hypothesis, with a distribution for the errors given by  $\widetilde F(.;  \mathbf{Y})$. In most cases,  $\Gamma(\widetilde F(.;  \mathbf{Y}))$ is invariant to the values of the fixed effects (see Remark \ref{Remark: invariance} for further details), so we can consider a DGP setting the fixed effects to zero. Therefore,  if the null is $H_0:\beta=0$, we  generate datasets with $\mathbf{Y}^b = \boldsymbol{\epsilon}^b$, where $\boldsymbol{\epsilon}^b$ is drawn from $\widetilde F(.;  \mathbf{Y})$. Using this simulated data, we calculate the proportion of times in which we would reject the null with this specific inference method, in a setting in which we know by construction that the null is true. This procedure is summarized in  the following algorithm:

	\begin{itemize}
	
	\item Step 1: do $b=1,...,B$ iterations of this step. In each step $b$:
	
		\begin{itemize}
		
		\item Step 1.1: draw a realization for $\boldsymbol{\epsilon}$ from  $\widetilde F(.;  \mathbf{Y})$, $\boldsymbol{\epsilon}^b$;
		
		\item Step 1.2: generate the outcomes $\mathbf{Y}^b = \boldsymbol{\epsilon}^b$ (see Remark \ref{Remark: invariance});
		
		\item Step 1.3: get $\hat \beta^b$ from  a regression of $\mathbf{Y}^b$ on $D_{it}$ and fixed effects;
		
		\item Step 1.4: test the null that the ATT equals zero using the inference method that is being assessed at a significance level $\alpha$. 
		
		\end{itemize}

	\item Step 2: $\widehat \Gamma(\widetilde F(.;  \mathbf{Y}))$ is the proportion of simulations in which the null is rejected. 
	
	\end{itemize}

Note that increasing the number of simulations $B$, $\widehat \Gamma(\widetilde F(.;  \mathbf{Y}))$ converges in probability to $ \Gamma(\widetilde F(.;  \mathbf{Y}))$ --- where  this convergence in probability is with respect to the uncertainty in the simulations, and it is conditional on $\mathbf{Y}$. Therefore, by choosing a large $B$ we can control for the simulation error. Given that, in what follows we use $ \Gamma(\widetilde F(.;  \mathbf{Y}))$, with the understanding that we can get a good approximation for this value using a large number of simulations. Unless otherwise noted, we consider assessments with a significance level  $\alpha = 0.05$. While in most cases the choice of $\alpha$ does not alter the conclusions of the assessments, we discuss in Section \ref{Choosing_alpha} settings in which it would be important to construct assessments for different significance levels.

The key question then is how to choose $\widetilde F(.;  \mathbf{Y})$. We consider in Section \ref{Sec: DiD illustrations} different alternatives that applied researchers can use, highlighting their advantages and disadvantages.\footnote{In principle, one could calculate  $\sup_{F(.) \in \mathcal{F}} \Gamma(F(.))$, which would give the finite-sample size of the test relative to this class of distributions $\mathcal{F}$. We do not follow this approach for a series of reasons. 
First, it is not common for applied researchers to explicitly specify a family $\mathcal{F}$ in their applications, so in many cases it would not be obvious to applied researchers how to proceed. Second, as discussed in Section \ref{Sec: DID setting}, even if a family ${\mathcal{F}}$  is specified, the problem might be that it is too restrictive to account for  relevant features of the empirical application. In such cases, calculating $\sup_{F(.) \in \mathcal{F}} \Gamma(F(.))$ would not be informative about potential inference problems, while an assessment $\Gamma(\widetilde F(.;  \mathbf{Y}))$ might be informative.  Finally, calculating  $\sup_{F(.) \in \mathcal{F}} \Gamma(F(.))$ might be unfeasible or require high computational costs in practice, being an empirical-application-specific endeavor. Instead, we want to propose assessments that would be easy and low cost for applied researchers to implement, while still being informative. }

\begin{remark}
\label{Remark: invariance}
Consider a general linear model $\mathbf{Y} = \mathbf{X}\boldsymbol{\gamma} + \boldsymbol{\eta}$, where $\mathbf{Y}$ is an $N \times 1$ vector of outcomes, $\mathbf{X}$ is an $N \times K$ matrix of covariates, $\boldsymbol{\gamma}$ is an $K \times 1$ vector of parameters, and $\boldsymbol{\eta}$ is an $N \times 1$ vector of errors. Let the null hypothesis be $\mathbf{R} \boldsymbol{\gamma} = \mathbf{q}$, for a $J \times K$ matrix $\mathbf{R}$ and a $J \times 1$ vector $\mathbf{q}$. We consider assessments based on simulations with $\mathbf{Y}^b = \mathbf{X}\widetilde{\boldsymbol{\gamma}} + \boldsymbol{\eta}^b$, where $\widetilde{\boldsymbol{\gamma}}$ satisfies the null and  $\boldsymbol{\eta}^b$ is drawn from a distribution $\widetilde F(.,\mathbf{Y})$. In this case, the OLS estimator of $\mathbf{Y}^b$ on $\mathbf{X}$ is such that $\mathbf{R} \widehat{\boldsymbol{\gamma}}^b - \mathbf{q} = \mathbf{R} (\mathbf{X}'\mathbf{X})^{-1} \mathbf{X}' \boldsymbol{\eta}^b $, while the residuals in the simulations  are $\widehat{\boldsymbol{\eta}}^b = \left( \mathbb{I} - \mathbf{X}(\mathbf{X}'\mathbf{X})^{-1} \mathbf{X}' \right) \boldsymbol{\eta}^b $. Therefore, any test such that the decision rule is a function of $\widehat{\boldsymbol{\gamma}}^b$  and $\widehat{\boldsymbol{\eta}}^b$ will be numerically invariant to the choice of $\widetilde{\boldsymbol{\gamma}}$ (provided it satisfies the null). In those cases, we can define $\Gamma(\widetilde F(.,\mathbf{Y}))$ without the need of specifying $\widetilde{\boldsymbol{\gamma}}$. Also, if we are assessing a test for a null in which one component of $\boldsymbol{\gamma}$ equals zero, then we can simply set $\widetilde{\boldsymbol{\gamma}}=0$, so $\mathbf{Y}^b = \boldsymbol{\eta}^b$. This is the reason why we can set the fixed effects to zero in the simulations for the DiD illustration above when we construct assessments for robust or cluster-robust standard errors. Moreover, since the ratio between $\widehat{\boldsymbol{\gamma}}^b$ and $\widehat{\boldsymbol{\eta}}^b$ is invariant to the scale of the distribution  $\widetilde F(.,\mathbf{Y})$, in most cases the assessment will also be numerically invariant to the \emph{scale} of the distribution  $\widetilde F(.,\mathbf{Y})$.

\end{remark}

\subsection{Inference assessments in different scenarios} \label{Sec: DiD illustrations}

We consider different scenarios in this DiD setting in which we can highlight advantages and disadvantages of assessments using different choices for $\widetilde F(.;  \mathbf{Y})$.

\subsubsection{Setting with $N_1=1$} \label{Sec_DID_N1=1}

We start considering a setting with $N_1=1$. This is a setting in which cluster-robust standard errors should perform very poorly as it relies on an asymptotic theory in which $N_1$ and $N_0$ are large. If errors are independent across states, then $\mathbb{V}(\hat \beta_{\mbox{\tiny DiD}}) = \mathbb{V}(W_1) + \frac{1}{N_0} \sum_{i=2}^N \mathbb{V}(W_i)$. However,  cluster-robust standard errors at the state level would estimate the component of the variance related to the treated, $\mathbb{V}(W_1)$, as zero (see details in \cite{AFW_Few_Treated}). Therefore, the estimated variance can be severely underestimated.  Still, cluster-robust standard errors have been widely used in DiD settings with one treated state even in  published papers.\footnote{This is the case, for example,  in a series of papers that estimate the effects of the Massachusetts 2006 health care reform \citep{Sommers,Miller,Niu,Courtemanche,KOLSTAD2012909}. For example, \cite{Sommers}  compared 14 Massachusetts counties with 513 control counties, and reported state-level cluster-robust standard errors. Their inference procedures were then revisited by \cite{Kaestner} and \cite{Ferman2020inference}. 
\label{Footnote_MA}} 

We analyze the properties of different types of assessments that applied researchers could run in this setting. We set $(N_1,N_0,T,t^\ast) = (1,50,20,10)$, while the \emph{true} distribution of the errors is AR(1) normal with correlation coefficient 0.3, and is iid across $i$.  

\textbf{Assessment based on iid normals:} a very simple assessment an applied researcher could run to check the reliability of inference using cluster-robust standard errors in this setting would be to set $\widetilde F(.;  \mathbf{Y}) = \Phi(.)$ as a standard normal iid distribution, and calculate $ \Gamma_{\mbox{\tiny cluster}}(\Phi(.))$. This is an easy assessment to implement, as the applied researcher would simply have to run simulations replacing the   original $\mathbf{Y}$ with a vector of iid standard normal variables.  As discussed in Remark \ref{Remark: invariance}, the assessment is invariant to the scale of the errors. Therefore, there is no loss in normalizing the variance to one. In this case, we would have $\Gamma_{\mbox{\tiny cluster}}(\Phi(.))=0.787$. Therefore, this assessment, without specifying more complex features of the errors, such as heteroskedasticity or serial correlations, would have provided an immediate (and correct) conclusion that clustering at the state level should be considered with caution.

Note that setting $\widetilde F(.;  \mathbf{Y}) = \Phi(.)$ does not mean that we believe   errors are homoskedastic normal, nor that there is no serial correlation. Still, the performance of cluster-robust standard errors even when we consider this simple DGP should indicate that it may also work poorly in potentially more complex distributions for the errors. Moreover, while the rejection rates might be lower for other distribution of the errors, if we consider a family of distributions $\mathcal{F}$ that includes iid normal errors, then  $ \Gamma_{\mbox{\tiny cluster}}(\Phi(.))$ would provide an easy-to-compute and very informative lower bound for the finite-sample size of the test relative to the family $\mathcal{F}$. That is, we know that  $\sup_{F(.) \in \mathcal{F}} \Gamma_{\mbox{\tiny cluster}}(F(.)) \geq \Gamma_{\mbox{\tiny cluster}}(\Phi(.)) = 0.787$. 

The fact that we consider a DGP without serial correlation does not prevent us from detecting that there is a problem with cluster-robust standard errors in this setting with one treated cluster. This happens because cluster-robust standard errors at the state level rely on an asymptotic theory in which $N_1$ and $N_0$ must be large, regardless of the within-cluster correlation. In particular, this is still true if there is no within-cluster correlation.

However, assessments based on $\Phi(.)$ would not be able to detect problems for inference methods that would work well in an iid normal setting, but that would perform poorly if the distribution of the error exhibits other features. For example, suppose we rely on robust standard errors (without cluster), and let  $\Gamma_{\mbox{\tiny robust}}(\Phi(.))$ be the assessment for this inference method using iid standard normals. In this case, $ \Gamma_{\mbox{\tiny robust}}(\Phi(.))=0.075$, while in the true DGP (that exhibits serial correlation), the rejection rate when relying on robust standard errors is higher, at 0.173. Since the DGP used in this assessment does not have serial correlation by construction, it is not able to detect that the over-rejection for the robust standard errors is much higher in this setting. 

\textbf{Assessment based on residual bootstrap:} another alternative would be to estimate the distribution of the errors using the residuals of the DiD regression. In this case, we would have a distribution for the assessment $\widetilde F(.;  \mathbf{Y})$ that is a function of the data. Let $\widehat{\boldsymbol{\epsilon}}_i$ be the $T \times 1$ vector of residuals for state $i$.  One alternative in this case is to construct a distribution $\widetilde F_{\mbox{\tiny rb}}(.;  \mathbf{Y})$ in which we sample with replacement vectors ${\boldsymbol{\epsilon}}^b_i$ from $\{\widehat{\boldsymbol{\epsilon}}_j\}_{j=1}^N$ for each state $i$. Since this is similar in spirit to a residual bootstrap at the state level, we call that an assessment based on a residual bootstrap, where  $\Gamma_{\mbox{\tiny robust}}(\widetilde F_{\mbox{\tiny rb}}(.;  \mathbf{Y}))$ applies this estimated DGP to assess inference based on robust standard errors. 

The main advantage of this assessment is that now we consider a distribution that  allows for serial correlation. In particular, if errors $\epsilon_i$ are iid across $i$, then $\widetilde F_{\mbox{\tiny rb}}(.;  \mathbf{Y})$ would asymptotically recover the true distribution of the errors when $N \rightarrow \infty$.

Differently from $ \Gamma_{\mbox{\tiny robust}}(\Phi(.))$, we would now get different assessments depending on the realization of the data, $\mathbf{Y}$. In order to evaluate how $\Gamma_{\mbox{\tiny robust}}(\widetilde F_{\mbox{\tiny rb}}(.;  \mathbf{Y}))$ varies with the realization of $\mathbf{Y}$, we consider 5,000 realizations of $\mathbf{Y}$ drawn from the \emph{true} distribution of $\boldsymbol{\epsilon}$, and then for each realization we calculate $\widehat \Gamma_{\mbox{\tiny robust}}(\widetilde F_{\mbox{\tiny rb}}(.;  \mathbf{Y}))$ using 1,000 simulations. This way, we recover the joint distribution of $(\mathbf{Y},\Gamma_{\mbox{\tiny robust}}(\widetilde F_{\mbox{\tiny rb}}(.;  \mathbf{Y})))$.

We present results for this assessment in Table \ref{Table_DiD_N1=1}. We find that in $93\%$ of the realizations of $\mathbf{Y}$, we would have an assessment $\Gamma_{\mbox{\tiny robust}}(\widetilde F_{\mbox{\tiny rb}}(.;  \mathbf{Y}))$ greater than $0.1$,  which would indicate that the use of robust standard errors should be used with caution.\footnote{The threshold of 0.1 to determine that the assessment would suggest problems with the inference method is admittedly arbitrary. Throughout the paper we would find similar results if we consider other thresholds.} In this case, $\Gamma_{\mbox{\tiny robust}}(\widetilde F_{\mbox{\tiny rb}}(.;  \mathbf{Y}))$ would (correctly) provide evidence that considering a family $\mathcal{F}$ that does not allow for serial correlation --- which is implicitly assumed when considering robust standard errors --- is too restrictive for this application, leading applied researchers to consider other alternatives for inference. In contrast, $\Gamma_{\mbox{\tiny robust}}(\Phi(.))$ would not detect this problem because the distribution $\Phi(.)$ would belong this family $\mathcal{F}$.

We also find that $\Gamma_{\mbox{\tiny cluster}}(\widetilde F_{\mbox{\tiny rb}}(.;  \mathbf{Y}))$ is always larger than $0.622$, so this assessment would also be effective in detecting that cluster-robust standard errors should be used with caution in this setting.

\textbf{Assessment based on non-parametric bootstrap:} while $\widetilde F_{\mbox{\tiny rb}}(.;  \mathbf{Y})$ can capture features of the distribution of the errors that $\Phi(.)$ does not capture, it  still imposes some restrictions, such as iid errors across states. Another alternative would be an assessment in the spirit of a non-parametric block-bootstrap conditional on the treatment assignment. In this case, we consider a DGP  $\widetilde F_{\mbox{\tiny npb}}(.;  \mathbf{Y})$ sampling with replacement the vector of errors for the treated (controls) from the residuals of the treated (controls). If we had many treated and many controls, and errors are independent across states, then  this approach would \emph{asymptotically} recover the distribution of the errors, allowing not only for serial correlation, but also for  different distributions for treated and control states. 

In this example, however, we have that $\Gamma_{\mbox{\tiny cluster}}(\widetilde F_{\mbox{\tiny npb}}(.;  \mathbf{Y}))$ is always around 0.05, and never higher than 0.10. Therefore, this assessment would fail to detect a problem that even a simple assessment with iid normals would be able to detect. This happens because we have only one treated state, so $\widetilde F_{\mbox{\tiny npb}}(.;  \mathbf{Y})$ recovers a distribution for the errors such that the variance of  $W_1$ equals zero. This provides an extreme example highlighting  that, while more complex assessments that use the data to estimate the DGP may have good asymptotic properties, they may not perform well in finite samples.  

\textbf{Discussion:} in this setting, if  applied researchers used even simple assessments such as $\Gamma_{\mbox{\tiny cluster}}(\Phi(.))$ or $\Gamma_{\mbox{\tiny cluster}}(\widetilde F_{\mbox{\tiny rb}}(.;  \mathbf{Y}))$, they would receive a strong warning that  inference based on cluster-robust standard errors at the state level should be used with caution. Given that, they  should analyze their empirical setting more carefully and consider a more appropriate inference method. 

In this particular example, researchers should figure out that cluster-robust standard errors at the state level is not valid when we have a single treated cluster, and should consider existing alternatives that do not rely on a large number of treated states (see \cite{AFW_Few_Treated} for a survey on inference methods with few treated units). Importantly, those alternatives rely on different sets of assumptions, and applied researchers should evaluate whether these alternative assumptions are reasonable in their applications. 

Considering the example of the scientific evidence on the effects of the Massachusetts 2006 health care reform (see Footnote \ref{Footnote_MA}), we see a series of papers that were published using cluster-robust standard errors with one treated cluster years after this problem  has been documented in the econometrics literature (which happened at least since \cite{Conley2011}). This reveals a lag from the time in which inference problems are uncovered, and the widespread knowledge of these conclusions to applied researchers/editors/referees, highlighting that inference assessments may remain relevant even after such econometrics papers have been published. Alternatively, it may be that assessments reveal inference problems that have been overlooked in the econometrics literature, so this may trigger new econometrics papers that analyze these problems and propose novel alternatives. 

We also emphasize that an assessment close to 5\% is not a guarantee that the inference method is reliable in a specific application. As we see in this example, $\Gamma_{\mbox{\tiny robust}}(\Phi(.))$ is relatively close to 5\%, because it is unable to consider that serial correlation invalidates the use of robust standard errors. We also see one case in which $\Gamma_{\mbox{\tiny cluster}}(\widetilde F_{\mbox{\tiny npb}}(.;  \mathbf{Y}))$ is close to 5\% because $\widetilde F_{\mbox{\tiny npb}}(.;  \mathbf{Y})$ recovers an unreasonable distribution for the errors, given that we are in a setting with few treated states (which is exactly a setting in which  we would like that assessments detect problems).  As another example, even in settings with many treated and many control states, none of these  assessments would  allow, by construction, for spatial correlation, which might be relevant in DiD applications \citep{Ferman_DID}.

Therefore, when an assessment is close to 5\%, it is crucial that applied researchers understand the limitations of the assessments they are using. In particular, they should understand what kind of problems an assessment would not be able to detect, given the choice for $\widetilde F(.;\mathbf{Y})$.

\subsubsection{Setting with $N_1=20$ and $N_0 = 180$; non-normal errors}
 \label{Sec: DiD lognormal}
Another example in which an inference assessment based on iid normals may fail to detect relevant problems is when errors are not normally distributed. In such cases, it may be that asymptotic approximations are less accurate for a given sample size than it would be if errors were normal.  To illustrate this issue, we consider the use of different types of assessments in a setting with  $(N_1,N_0)=(20,180)$, while errors $\boldsymbol{\epsilon}_i$ are such that $W_i$ has a log-normal distribution with the mean normalized to zero for the treated and standard normal distribution for the controls. Given this distribution for the errors, the true size of the test based on cluster-robust standard errors is $14\%$.\footnote{In this setting, the DiD estimator and the cluster-robust standard errors at the state level are numerically equivalent to a treatment-control comparison of post-pre differences using robust standard errors. Therefore, we consider simulations already using the collapsed post-pre data, without the need of specifying details such as $T$, $t^\ast$, and the distribution of $\epsilon_{it}$.} The results of different types of assessments in this setting are presented in Table \ref{Table_lognormal}.

\textbf{Assessment based on iid normals:} first,  note that $\Gamma_{\mbox{\tiny cluster}} (\Phi(.)) = 7\%$, understating the over-rejection problem due to the fact that the errors are not normally distributed for the treated. Not surprisingly, one of the main limitations of this assessment is that it does not take into account the possibility that errors are not normally distributed. 

\textbf{Assessment based on residual bootstrap:}
$\Gamma_{\mbox{\tiny cluster}}(\widetilde F_{\mbox{\tiny rb}}(.;  \mathbf{Y}))$ in this case is also close to 0.05, being smaller than 0.10 for all realizations of $\mathbf{Y}$. Therefore, even though this assessment uses the data to estimate the distribution of the errors, it also fails to detect that asymptotic approximations are less reliable because errors are not normally distributed. 

\textbf{Assessment based on non-parametric bootstrap:} we consider then $\Gamma_{\mbox{\tiny cluster}}(\widetilde F_{\mbox{\tiny npb}}(.;  \mathbf{Y}))$. This assessment would indicate that the inference method might present relevant size distortions with a  positive probability. For example, in  49\% of the realizations of $\mathbf{Y}$ the assessment  $\Gamma_{\mbox{\tiny cluster}}(\widetilde F_{\mbox{\tiny npb}}(.;  \mathbf{Y}))$ would be greater than 0.1.

However, an important issue in this case is that, in approximately half of the realizations of the data, the assessment would fail to detect large distortions, which could not trigger applied researchers to review the inference methods they are using. Since for those realizations of the data applied researchers would continue to rely on such inference method, this may create  sequential-testing problems.\footnote{A non-exhaustive list of papers that analyze the implications of pre-testing in different settings include \cite{10.1162/REST_a_00682}, \cite{10.2307/40664469}, and \cite{Roth}. To the best of our knowledge, this issue has not been discussed in the context of inference assessments.} More specifically, if we define a threshold $\kappa$ such that applied researchers would continue to rely on this inference methods if $\Gamma_{\mbox{\tiny cluster}}(\widetilde F_{\mbox{\tiny npb}}(.;  \mathbf{Y})) < \kappa$, then we should consider the properties of the estimator conditional on that. In this example, we have that $\mathbb{E}[\hat \beta |\Gamma_{\mbox{\tiny cluster}}(\widetilde F_{\mbox{\tiny npb}}(.;  \mathbf{Y})) < 0.1 ] \neq  \beta =0$. Therefore, even though the estimator $\hat \beta$ is unbiased, this sequential-testing procedure introduces bias once we condition on the assessment failing to detect the problem with the inference method. 

\textbf{Assessment based on WGAN:} we can also consider the use of Wasserstein Generative Adversarial Networks (WGAN), which allows for more flexible approximations for $F(\boldsymbol{\epsilon})$ \citep{arjovsky2017wasserstein,athey2020using}. Implementing the WGAN involves a number of implementation decisions and tuning parameters. We consider two different implementations, varying the batch size and maximum number of epochs, which can be either $(50,2000)$ or $(100,500)$.\footnote{We have downloaded the Python package from \href{https://github.com/gsbDBI/ds-wgan/}{https://github.com/gsbDBI/ds-wgan/}, and mainly used the package’s standard parameters in our implementation.}  In the first implementation, $\Gamma_{\mbox{\tiny cluster}}(\widetilde F_{\mbox{\tiny WGAN1}}(.;  \mathbf{Y}))$  would be greater than 0.10 in 35\% of the realizations of $\mathbf{Y}$, which would raise a red flag that inference may be unreliable. However, there would still be a large probability that the assessment is smaller than 0.1, and the estimator is biased if we condition on that. In contrast, for the alternative implementation of the WGAN, we never find assessments greater than 0.1 across realizations of $\mathbf{Y}$. 

Overall, these results show that, for some implementation choices, the use of WGAN can capture features of the distribution of the errors  that other choices for $\widetilde{F}(.;\mathbf{Y})$, such as $\Phi(.)$ or $\widetilde{F}_{\mbox{\tiny rb}}(.;\mathbf{Y})$, would not capture. However, since it depends on the  estimated residuals, this type of assessment is subject to sequential-testing problems. This problem appears exactly for the WGAN implementation that detects size distortions due to the log-normal distribution of the errors with a positive probability. Moreover, this method requires a number of choices on tuning parameters, and we show that different choices may lead to different conclusions. Therefore, applied researchers would be able to cherry pick tuning parameters in which the assessment would not suggest much distortions. Finally, we note that implementing a WGAN is substantially more complex than implementing the other assessments we discuss. Therefore, implementing this type of assessment would arguably  be more costly for applied researchers. 

\textbf{Discussion:} we show that considering more complex assessments that use data to estimate a distribution for the errors may be helpful in terms of detecting features of this distribution that would not be captured by a simpler assessment based on iid normals. However, a potential drawback in relying in such assessments is that they are subject to sequential-testing distortions. In this example, the assessments that are able to detect larger over-rejections are also those that exhibit sequential-testing distortions. Other drawbacks include higher implementation costs, and excess of degrees-of-freedom in determining $\widetilde F(.;\mathbf{Y})$, which may create opportunity for cherry picking assessments that would not indicate problems with the inference method.

\subsubsection{Setting with large \texorpdfstring{$N_1$}{N1} and large \texorpdfstring{$N_0$}{N0}}

 \label{Sec: problems even with large N}

Inference assessments may also be relevant  in settings in which the number of treated and control states are very large. This may be the case  when there is an implementation problem in the calculation of the standard errors, so they are invalid even asymptotically.

In the DiD setting, this may happen, for example, if  applied researchers do not use the appropriate degrees-of-freedom correction when calculating the standard errors. As discussed by \cite{cameron2015practitioner}, if we are clustering at the state level, then we should not take into account the number of state fixed effects in the degrees-of-freedom correction. However, depending on the command used in Stata or R to calculate the standard errors, it will (incorrectly) take that into account.

In order to illustrate that, consider a DiD setting with $(N_1,N_0,T,t^\ast)=(250,250,4,2)$. In this case, the cluster-robust standard errors will always be around 15\% \emph{larger} when we use the incorrect  degrees-of-freedom correction, regardless of the distribution of the errors. Therefore, even the assessment with iid normals would reveal a rejection rate of $0.025\%$, which is \emph{lower} than the nominal size of the test, correctly indicating that there is a problem in the calculation of the standard errors.  

Interestingly, this is a setting in which the inference method presents a problem regardless of the distribution of the errors. Therefore, considering iid normal errors is sufficient to detect such problems, despite the fact that it does not approximate the true DGP even asymptotically. Another illustration in which cluster-robust standard errors may be asymptotically incorrect due to issues with degrees-of-freedom correction is when we consider experiments with pairwise stratification of schools, and we cluster the standard errors at the school level in student-level regressions. As documented by \cite{Chaisemartin_cluster}, standard errors in this case would be underestimated, leading to over-rejection. Again, even simple assessments with iid normals would easily reveal this problem.

\subsubsection{Comparison with \cite{Advani}  \& setting with $N_1=N_0=15$} \label{Sec: DID comparison with Advani}

Overall, our conclusions regarding the use of simulations to assess inference methods differ from the more skeptical views from \cite{Advani} regarding the use of Empirical Monte Carlo studies (EMCS) for estimator selection. \cite{Advani} consider a setting in which applied researchers want to rank a set of  estimators under unconfoundness, based on finite-sample properties such as bias or mean squared errors. They analyze whether EMCS would be able to correctly rank those estimators, reaching  pessimistic conclusions. 

In contrast, our proposal is \emph{not} that applied researchers start with a set of inference methods, and then choose the one with smallest size distortion in a pre-specified assessment. Rather, the idea is that such assessments may be informative for applied researchers in case the inference method they were intending to use is not appropriate for their empirical settings. For example, in the illustration from Section \ref{Sec_DID_N1=1}  researchers would have figured out that state-level cluster is unreliable in this setting with one treated cluster. Given that, they should consider alternative inference methods, being mindful about the assumptions that such alternatives rely on. 

Consider another example to illustrate this difference relative to the conclusions from \cite{Advani}. Let $(N_1,N_0,T,t^\ast) = (15,15,20,10)$. In such setting, we have that $\Gamma_{\mbox{\tiny cluster}}(\Phi(.)) = 0.06$, while $\Gamma_{\mbox{\tiny robust}}(\Phi(.)) = 0.05$. Therefore, the assessment with iid normals would indicate a slight over-rejection for inference based on cluster-robust standard errors, while the assessment for robust standard errors would be closer to 5\%. Therefore, if we were simply choosing among these options based on the inference assessment, we would choose inference based on robust standard errors.

Importantly, however, we are \emph{not} recommending that applied researchers  start with a set of inference methods (in this case, inference based on cluster-robust standard errors or on robust standard errors) and choose among these options based on the results of a pre-specified assessment (in this case, an assessment based on iid normals). Instead of simply choosing robust standard errors (given an assessment closer to 5\%), applied researchers should have an understanding about the assumptions that each of those methods rely on, and about the type of problems that the assessment would be able to detect. In this case, it should be clear that robust standard errors do not allow for serial correlation, and that an iid-normal assessment would not detect problems due to serial correlation.

\section{ Shift-Share Designs: Design-based Assessments}
\label{Sec: shift-share}

\subsection{setting} \label{Sec: SS setting}

We consider now shift-share designs, which are specifications that study the impact of a set of shocks on units differentially exposed to them. Consider a setting with $i=1,...,N$ regions that are subject to $f=1,...,F$ aggregate shocks $\mathcal{X}_f$. The shift-share variable is given by ${x}_i = \sum_{f=1}^F w_{if}\mathcal{X}_f$, where the shares $w_{if}$ reflect how shock $\mathcal{X}_f$ affects unit $i$. The regression model is then given by
\begin{eqnarray}
y_i = \beta_0 + \beta x_i + \eta_i,
\end{eqnarray}
where $\beta$ reflects the effect of $x_i$ on $y_i$. Let $\mathbf{y}$ be the $N \times 1$ vector of outcomes, and $\Omega$ be the $N\times F$ matrix with all shares for all regions.  

As highlighted by \cite{Adao}, if we want to test the null $H_0: \beta=0$, an important challenge in this setting is that the error term $\eta_i$ may be spatially correlated. In particular,  if regions with similar shares tend to have correlated errors, this spatial correlation leads to over-rejection if one relies on robust or cluster-robust standard errors.   In order to illustrate this issue, \cite{Adao} consider simulations in which they hold  $\mathbf{y}$ and $\Omega$ fixed, and draw realizations of $\{\tilde{\mathcal{X}}_f\}_{f=1}^F$ for a given distribution for the shocks. For each realization of the shocks in these simulations, they estimate the shift-share estimator from  $y_i = \gamma_0 + \gamma \tilde x_i^b + \tilde \eta_i$, where $\tilde x_i^b = \sum_{f=1}^F w_{if}\tilde{\mathcal{X}}^b_f$, and test the null $\gamma=0$ using robust or cluster-robust standard errors. In these simulations, they find rejection rates much higher than the nominal size of the tests. We call this type of simulations in which the stochastic component comes from the realization of the covariates as design-based assessments, and  discuss its  properties in Section \ref{Sec: design-based assessment}.     

\cite{Adao} and \cite{Borusyak} propose interesting alternatives for inference taking into account the possibility of spatial correlation. They consider a 
design-based setting, where shocks $\mathcal{X}_f$ are stochastic, while potential outcomes and shares are conditioned on.\footnote{See \cite{GP} for an alternative sampling framework for shift-share designs.}
A simplified version of their potential outcomes model is given by $y_i(x) = y_i(0) + \beta x$, where we assume for simplicity linearity and treatment effect homogeneity across shocks. In this setting, the identification assumption is that $\mathbb{E}[\mathcal{X}_f | \mathcal{L}]=c$ for a constant $c$, where $\mathcal{L}$ includes  potential outcomes and  shares. The inference methods they developed  are asymptotically valid in this sampling framework when (i) shocks are independent, (ii) the number of shocks goes to infinity, and (iii) the size of each shock becomes asymptotically negligible.\footnote{They also allow for clusters of shocks. In this case, the assumption is that shocks from different clusters are independent, the number of clusters of shocks goes to infinity, and  the size of each cluster of shocks  becomes asymptotically negligible.} The fact that they developed their theory in this framework does not necessarily mean that the focus in shift-share designs should be on inference conditional on $\mathcal{L}$. As discussed by \cite{Adao}, the idea of conditioning on potential outcomes is so that they ``can allow for any correlation structure of the regression residuals across regions.''

\subsection{Design-based Assessments}
\label{Sec: design-based assessment}

\subsubsection{Design-based assessments in randomized controlled trials}

For ease of exposition, we start discussing the use of design-based assessments in a randomized controlled trial (RCT). As we then show in Section \ref{Sec: DB in SS}, some of the insights from this simpler case of an RCT can be extrapolated to shift-share designs.

Consider an RCT in which we observe $i=1,...,N$ individuals whose potential outcomes are given by $Y_i(0)$ and $Y_i(1)$. We consider a design-based setting in which  potential outcomes are fixed, so the only source of uncertainty comes from the treatment allocation $\mathbf{T} = (T_1,...,T_N)$, which is assumed to have a known distribution \citep{Finite_pop,NBERw24003,fisher,Neyman1990}. The target parameter is the sample average treatment effect  (SATE), $\frac{1}{N} \sum_{i=1}^N (Y_i(1) - Y_i(0))$.\footnote{\cite{Finite_pop} and \cite{NBERw24003} also consider the possibility of sampling uncertainty. In this case, an alternative target parameter would be the population average treatment effect. For simplicity, we abstract from that.} In this case, inference using robust standard errors is asymptotically valid for the SATE in this sampling framework when $N \rightarrow \infty$ (although in some cases it may be conservative). 

Interestingly, if we knew $Y_i(0)$ and $Y_i(1)$ for all $i$, then knowledge of the distribution of $\mathbf{T}$ would provide all information needed to recover the true probability of rejection for a test based on, for example, robust standard errors. More specifically, note that the t-statistic with robust standard errors will be a function of $\{Y_i(0),Y_i(1)\}_{i=1}^N$ and $\mathbf{T}$, where the only stochastic component in this design-based setting is $\mathbf{T}$. Therefore, we would be able to calculate the exact probability that the t-statistic would be greater than the critical value and, thus, the probability of rejection. However, we do not know $Y_i(0)$ and $Y_i(1)$ for all $i$.  

A commonly used way to construct an assessment in this setting is to hold the vector of realized outcomes $\mathbf{Y}$ fixed, and consider different allocations of $\mathbf{T}$. Then, for each allocation of $\mathbf{T}$ we estimate the parameter of interest and test the null using the inference method being assessed.\footnote{Note that this type of design-based assessment is conceptually different from using randomization inference in RCT's. In randomization inference, we use these permutations to test a sharp null hypothesis that $Y_i(1) = Y_i(0)$ for all $i$. In contrast, this design-based assessment  uses the permutations to assess whether an inference method (for example, t-test with robust standard errors) is reliable for testing an hypothesis regarding the SATE.} This design-based assessment considers a DGP in which potential outcomes are given by  $\{ \widetilde Y_i(0),\widetilde Y_i(1) \}_{i=1}^N$, where $\widetilde Y_i(1) = \widetilde Y_i(0) = Y_i$. Therefore, in this DGP the null hypothesis that the SATE equals zero is true, because $\frac{1}{N} \sum_{i=1}^N (\widetilde Y_i(1) - \widetilde Y_i(0))=0$, and the design-based assessment gives us the size of this test if potential outcomes were given by $\{ \widetilde Y_i(0),\widetilde Y_i(1) \}_{i=1}^N$. 

However, the true DGP would not  be given by $\{ \widetilde Y_i(0),\widetilde Y_i(1) \}_{i=1}^N$ if  $Y_i(1) \neq Y_i(0)$ for some $i$. Therefore, even if we know the distribution of $\mathbf{T}$, a design-based assessment would not necessarily recover the true rejection rate for a given inference method. As we show in Section \ref{Sec: DB in SS}, this feature implies an important challenge for the type of simulations considered by \cite{Adao}, as the DGP used in the simulations might incorporate features that are not present in the true DGP.

\subsubsection{Design-based assessments in shift-share designs} \label{Sec: DB in SS}

Consider now the shift-share design setting from Section \ref{Sec: SS setting}, and suppose an applied researcher wants to assess whether inference with robust standard errors is reliable.  We  formalize what a design-based assessment as the one considered by \cite{Adao} would recover in this setting, depending on the true treatment effect $\beta$ and on the presence or absence of spatial correlation.

To this end,  consider a simpler version of the shift-share design setting from Section \ref{Sec: SS setting} in which observations $i=1,...,N$ are partitioned into equally-sized groups $\Lambda_1,...,\Lambda_F$, with $w_{if} = 1$ if $i \in \Lambda_f$, and $w_{if} = 0$ otherwise. Assume also that $\mathcal{X}_f \in \{0,1\}$ and $ \sum_{f=1}^F \mathcal{X}_f =F/2$. This particular shift-share design setting can be seen as an RCT in which treatment is assigned at the group $\Lambda_f$ level.
In this setting, we know that inference based on robust standard errors would be asymptotically valid when $N \rightarrow \infty$ if errors are independent across $i$. However, 
it would lead to over-rejection in case errors within partitions are  positive correlated.

We consider an inference assessment with random draws of $\widetilde{\mathcal{X}}_f \in \{0,1\}$ such that $\sum_{f=1}^F \widetilde{\mathcal{X}}_f =F/2$, while holding $\mathbf{y}$ fixed.  More specifically, for each draw of $\widetilde{\mathcal{X}}_f \in \{0,1\}$, we run the regression of $y_i = \gamma_0 + \gamma \tilde x_i + \tilde \eta_i$, where $\tilde x_i = \sum_{f=1}^F w_{if} \tilde{\mathcal{X}_f}$, yielding $\hat \gamma^b$. Then we test the null $\gamma=0$ using robust standard errors at a significance level $\alpha$, and the assessment is the rejection rate in these simulations. The DGP in the simulations sets  potential outcomes as $\tilde y_i(0) = \tilde y_i(1) = y_i$ (which are fixed, given the sampling framework of the simulations) and the distribution for $\widetilde{\mathcal{X}}_f $ described above.\footnote{Given the structure of this shift-share design example, we only need to define the potential outcomes $y_i(x)$ for $x \in \{0,1\}$.} Therefore, the null hypothesis $H_0:\gamma=0$ is true in this DGP.

Uncertainty in these simulations comes only from realizations of $ \widetilde{\mathcal{X}}_f$. Let $\mathbb{E}^\ast[.| \mathbf{y}]$ and $\mathbb{V}^\ast(. | \mathbf{y})$ denote the expectation and variance operators with respect to  this measure, conditional on $\mathbf{y}$. From Lemma 5 from \cite{IK}, $\mathbb{E}^\ast [\hat \gamma^b | \mathbf{y} ] =0$, so  the estimator $\hat \gamma^b$ is unbiased. Let $\mathbb{V}^\ast_{\mbox{\tiny true}} \equiv \mathbb{V}^\ast\left(\hat \gamma^b | \mathbf{y} \right)$ be the true variance of $\hat \gamma^b$ in these design-based simulations. Note that this is a number for a fixed $\mathbf{y}$, and a random variable depending on the errors $\epsilon_i$ when $\mathbf{y}$ is treated as a random variable. Also, let $\mathbb{V}^\ast_{\mbox{\tiny robust}}$ be the true variance in case treatment were assigned at the individual level in the design-based simulation. This is what the robust standard errors would asymptotically recover in these simulations when $N \rightarrow \infty$. 

The ratio ${\mathbb{V}^\ast_{\mbox{\tiny robust}}}/{\mathbb{V}^\ast_{\mbox{\tiny true}}}$ is crucial to understand whether the design-based simulations would indicate or not relevant over-rejections when we assess inference using robust standard errors. Considering this framework with $
\mathbf{y}$ fixed, if ${\mathbb{V}^\ast_{\mbox{\tiny robust}}}/{\mathbb{V}^\ast_{\mbox{\tiny true}}}$ converges to a value smaller than one in a sequence with $N \rightarrow \infty$, this means that the robust variance estimator  in the simulations would tend to underestimate the true variance in the simulations, generating an assessment larger than $\alpha$. In contrast, if ${\mathbb{V}^\ast_{\mbox{\tiny robust}}}/{\mathbb{V}^\ast_{\mbox{\tiny true}}}$ converges to one in a sequence with $N \rightarrow \infty$, then the robust variance calculated in the simulations would adequately estimate the true variance in the simulations, so we should expect assessments close to $\alpha$.

In this setting, we have the following result.

\begin{proposition} \label{Prop}
     Consider the shift-share design setting described in this section, and assume the vectors $\{\epsilon_i: i \in \Lambda_f\}$ are iid across $f$ with $\mathbb{E}[\epsilon_i]=0$, $\mathbb{V}(\epsilon_i)=\sigma^2$, and $cov(\epsilon_i,\epsilon_s)=\rho$ for $i \neq s$ and $i,s \in \Lambda_f$ for some $f$. Consider an asymptotic sequence in which $F \rightarrow \infty$ where we maintain that each group $\Lambda_f$ has exactly $m$ observations (so $N = m \times F$) and $\sum_{f=1}^F \mathcal{X}_f=F/2$. Then 
    \begin{eqnarray}
        \frac{\mathbb{V}^\ast_{\mbox{\tiny robust}}}{\mathbb{V}^\ast_{\mbox{\tiny true}}} \overset{a.s.}{\rightarrow} \frac{\beta^2 + 4 \sigma^2}{m\beta^2 + 4 \sigma^2 + 4(m-1)\rho}.
    \end{eqnarray}
\end{proposition}

\begin{proof}
    See Appendix \ref{Appendix_design_based}.
\end{proof}

First, consider a setting with no spatial correlation, so $\rho =0$. This is a setting in which robust standard errors would be asymptotically valid when $N \rightarrow \infty$. However, Proposition \ref{Prop} shows that, with probability one, ${\mathbb{V}^\ast_{\mbox{\tiny robust}}}/{\mathbb{V}^\ast_{\mbox{\tiny true}}} \rightarrow k<1$, whenever the true DGP exhibits a true treatment effect ($\beta \neq 0$) and $m>1$. Therefore, the design-based assessment would tend to be larger than $\alpha$, (incorrectly) suggesting that robust standard errors are invalid due to spatial correlation. The intuition is that the true treatment effect $\beta$ is confounded with spatial correlation that affects the $m>1$ units within the same partition in a similar way in the DGP used in the simulations.\footnote{The same conclusion is also valid in DiD settings if we want to assess whether robust standard errors are invalid due to serial correlation using a design-based assessment in a specific empirical application. This is not a problem for the simulations considered by \cite{Bertrand04howmuch}, since they did not applied these simulations to a specific empirical application.}  

Consider now a setting in which we should expect $\beta=0$ (for example, when we consider pre-treatment data to run the simulations). In this case, with probability one,  ${\mathbb{V}^\ast_{\mbox{\tiny robust}}}/{\mathbb{V}^\ast_{\mbox{\tiny true}}} \rightarrow 1$ when $\rho =0$. Therefore, if we are in a setting with no spatial correlation (and $N$ is large),  this  design-based assessment would be close to $\alpha$, (correctly) suggesting that robust standard errors would be asymptotically valid. In contrast, if we are in a setting with $\rho >0$, then the design-based assessment would be larger than $\alpha$, (correctly) suggesting that robust standard errors are invalid due to spatial correlation. Therefore, even though these simulations hold $\mathbf{y}$ fixed and consider only covariates as stochastic, this formalization also clarifies that, considering a setting with no treatment effect ($\beta=0$), these simulations are informative about potential problems in the errors, such as spatial correlation.  

An alternative to assess the relevance of spatial correlation when $\beta \neq 0$ is to consider $\boldsymbol{\epsilon}$-fixed design-based assessments. In this case, we consider a DGP in which we fix potential outcomes as $\dot y_i(0) = \dot y_i(1) = y_i - \hat \beta x_i$. If potential outcomes in the true model are given by  $y_i(x) = y_i +\beta x$ and $\hat \beta$ is a consistent estimator for $\beta$, then we would have simulations with a structure for the potential outcomes that is more similar to the true structure  in the application. This is how \cite{Borusyak} construct their simulations in their Appendix A.11. Interestingly, if treatment effects are heterogeneous or non-linear, then treatment effects would still generate spatially correlated errors in the design-based DGP, even after subtracting $\hat \beta x_i$ to construct the potential outcomes for the simulations. Therefore, these simulations would still be informative about problems with robust standard errors in shift-share design regressions due to heterogeneous treatment effects.

Appendix \ref{Appendix_permutation_shocks} presents simulations on the use of this kind of assessments to detect spatial correlation. These simulations confirm these novel insights  (i) that a design-based assessment holding $\mathbf{y}$ fixed may incorrectly indicate spatial correlation; (ii) that an $\boldsymbol{\epsilon}$-fixed design-based assessment does not present this problem; (iii) that $\mathbf{y}$-fixed design-based assessments can be informative about spatial correlation when $\beta =0$ (for example, when we consider an assessment based on a placebo specification); and (iv) that these assessments are informative about inference problems when treatment effects are  heterogeneous.

\subsection{Design-based assessments in shift-share  empirical applications} \label{Sec: SS illustrations}

We consider now the use of design-based assessments in  three shift-share design empirical applications, based on \cite{Autor}, \cite{Dix}, and \cite{Acemoglu}. 

\textbf{Assessing cluster-robust standard errors:} we consider first the use of cluster-robust standard errors in these applications. We know that cluster-robust standard errors might be problematic in this setting if (i) there is spatial correlation in the errors (beyond the clusters), and/or (ii) there are few clusters. We show in Appendix Table \ref{Table_SS_CRVE} that, in the absence of spatial correlation, asymptotic approximations for cluster-robust standard errors are relatively less reliable for the weighted OLS specifications in these applications. Since the focus in this section is to understand how different assessments may be used to detect spatial-correlation problems, we therefore focus on the unweighted specifications.   We emphasize, though, that the assessments we analyze in this section would be able to detect both of these problems. 

We consider first a design-based assessment with $\mathbf{y}$ fixed and stochastic shocks, which  is exactly what \cite{Adao} considered to illustrate the possibility of spatial correlation in the errors in shift-share designs. We find large rejection rates in columns 1, 3 and 5 of Table \ref{Table_SS}, ranging from 34\% to 70\%. However, as discussed in Section \ref{Sec: DB in SS}, a true treatment effect of ${x}_i$ may be confounded with spatially-correlated errors in these simulations. Therefore, we may find large assessments, even when errors are \emph{not} spatially correlated. 

We consider the two alternatives discussed in Section \ref{Sec: DB in SS} to properly assess whether spatial correlation is a problem for cluster-robust standard errors in these applications. All results are presented in Table \ref{Table_SS}. 
For the application from \cite{Autor}, we reinforce the conclusions from \cite{Adao} that spatial correlation leads to relevant over-rejection in this application, although we find evidence that their assessments indeed over-state the magnitude of the problem (consistent with the theoretical results discussed in Section \ref{Sec: DB in SS}).

For \cite{Acemoglu}, the $\boldsymbol{\epsilon}$-fixed design-based assessment remains large, but the design-based assessment using a placebo outcome is close to 5\%. This provides evidence that inference based on cluster-robust standard errors might be reasonable for testing a sharp null of no effect whatsoever, but that we may have relevant heterogeneous treatment effects. As we show in Appendix \ref{Appendix_SS_applications}, the assessment using the pre-treatment outcome would have a relatively high probability of flagging problems due to spatial correlation in this application.

Finally, for \cite{Dix}, both alternatives lead to assessments smaller than 5\%, providing some indication that spatial correlation is not a problem in this  application. An important caveat, however, is that those assessments would have a relatively lower probability of flagging a problem in case there is relevant spatial correlation in this application (details in Appendix \ref{Appendix_SS_applications}). An alternative in cases like that may be to run assessments for a number of pre-treatment outcomes, if available.

\textbf{Assessing new inference methods for shift-share designs:}  we also assess the inference methods proposed by \cite{Adao}. We find evidence that these inference methods work well in \cite{Autor}. However, for the other two applications the assessments suggest that these inference methods can lead to large distortions, with rejection rates up to 57\% (Table \ref{Table_SS}, Panel C). This is consistent with these applications having a smaller number of sectors (details in Table \ref{Table_SS}). Note that it is not a problem to consider a $\mathbf{y}$-fixed design-based assessment in this case, because we are evaluating inference methods that allow for spatial correlation. Therefore, even if a true treatment effect is confounded with spatial correlation in the simulations, this would not be a problem in this case.

\textbf{Choosing among alternative inference methods in shift-share designs:}  these results illustrate that it is not trivial to determine which inference methods are more reliable in shift-share designs. If we have evidence that the methods from \cite{Adao}/\cite{Borusyak} are reliable, then they should be preferred, as they impose less restrictions on the spatial correlation. Therefore, our recommendation is to start with a design-based assessment to assess whether inference based on their methods is reliable.  This is the case for the application from \cite{Autor}. 

In case the design-based assessment suggests these new inference methods are unreliable, then the next step should be to use one of the alternative assessments we propose (design-based assessment on a placebo outcome and/or $\boldsymbol{\epsilon}$-fixed design-based assessment) to evaluate whether inference based on cluster-robust standard errors would be reliable. These assessments would be informative about two potential problems with cluster-robust standard errors in this setting: (i) in case the number of clusters is not large enough, and (ii) in case it is unreasonable to assume that there is no relevant spatial correlation.\footnote{In order to distinguish between these two problems, one could also run an assessment with both errors and shocks stochastic, as we do in Appendix Table \ref{Table_SS_CRVE}. In this case, if we consider a DGP in which errors are independent, then the assessment would be informative about the first problem, but not about the second one. } In our empirical illustrations,  we find evidence that inference based on cluster-robust standard errors is relatively more reliable than inference based on these new methods for the applications from \cite{Acemoglu} (for testing a sharp null), and from \cite{Dix} (with the caveat that the assessments have lower probabilities of detecting problems in this application).
The \emph{placebo} specifications in these two applications provide further evidence on the conclusion that, in these applications, cluster-robust standard errors are relatively more reliable: in both cases, we would reject the null for their \emph{placebo} specifications if we consider \cite{Adao} standard errors, while we do not reject these nulls if we use cluster-robust standard errors (Appendix Table \ref{Table_appendix}, columns 4 and 6).
 
Importantly, if applied researchers consider a standard (fixed-$\mathbf{y}$) design-based simulation resampling shocks in these last two applications (instead of the alternatives we propose), they might incorrectly conclude that cluster-robust standard errors are less reliable than the new inference methods in these applications. 

In case we have evidence that none of these methods are reliable, then one would have to consider other alternatives. For example, \cite{Borusyak2} and \cite{Alvarez} consider the use of randomization inference tests for shift-share designs, and show conditions in which these tests are exact in finite samples even when we allow for unrestricted spatial correlation. However, these tests rely on relatively strong assumptions on the shock assignment mechanism (such as correct specification of the distribution of shocks or exchangeability) for this finite-sample validity. Applied researcher would then have to decide on whether those are reasonable assumptions in their settings.

\section{Implementation decisions for inference assessments}
\label{Sec: implementation}

\subsection{What should be treated as fixed/stochastic?} \label{Source_uncertainty}

In Sections \ref{Sec: DiD} and \ref{Sec: shift-share} we discuss two different types of assessments, one in which we fix covariates and uncertainty comes from the errors (model-based assessments) and one in which we fix outcomes and uncertainty comes from the covariates (design-based assessments). An important question then is which type of assessment one should consider in specific empirical applications.

In many applications, the decision on what is stochastic is essentially linked to the target parameter of interest. For example, consider an RCT. When we define the target parameter as the SATE, we are implicitly assuming that uncertainty comes only from treatment assignment. In contrast, if we consider as a target parameter the ATE for a larger population/superpopulation, then potential outcomes would also be stochastic \citep{Finite_pop,Miratrix,2019arXiv191201052D}. Ideally, applied researchers would first define their target parameters and the sources of uncertainty in their settings, and then proceed with inference assessments consistent with these definitions. Alternatively, in some applications we are assessing an inference method that was designed to work in a specific sampling framework. In this case, it is natural to design assessments consistent with that, as we do in Section \ref{Sec: shift-share} for shift-share design applications.

We also note that assessments conditional on either covariates or outcomes might also be informative about whether inference methods are reliable for unconditional inference. For example, suppose an assessment conditional on covariates suggests large over-rejection. In this case, we would generally need under-rejection conditional on other values for the covariates, or that the realized covariates are a very low probability event, so that, unconditionally, the test size is close to $\alpha$. Therefore, if a conditional assessment suggests relevant distortions, this should at least raise a red flag that there \emph{might} be relevant distortions when we consider unconditional inference as well. Likewise, as formalized in Section \ref{Sec: DB in SS}, design-based assessments can be informative about the correlation structure of the errors, even though they hold outcomes/errors fixed in the simulations.

It might also be that a conditional assessments would (correctly) suggest distortions for conditional inference, while inference is reliable unconditionally. In our view, conditional assessments in those cases should at the very least lead researchers to think more carefully about whether conditional or unconditional inference is more appropriate in their applications. To illustrate that, consider a DiD setting in which we observe $M_i$ individual-level observations for each  states $i$, where a single state is treated. Assume errors for all individuals and all periods are iid $N(0,1)$.  Conditional on the treatment assignment, if the treated state has fewer individuals than the control states, then the inference method proposed by \cite{Conley2011} would over-reject (see \cite{FP} and \cite{MW} for details). An inference assessment based on iid normal would reveal that the method would over-reject when we consider inference conditional on the treatment assignment. However, if all states have the same probability of being treated, then this inference method would have the correct size unconditionally. Our view is that the conditional assessment in this case should,  at the very least, lead applied researchers to carefully think about whether conditional or unconditional inference is more relevant in their settings. For this particular example, \cite{FP} and \cite{MW} argue that inference conditional on the treatment allocation is more appropriate. 

Overall, as we discuss further in Section \ref{Recommendations}, we see assessments conditional on covariates as appropriate and informative in any setting in which it is reasonable to think of potential outcomes as stochastic, at least as an initial screening.

\subsection{Which \texorpdfstring{$\alpha$}{alpha} to choose?} \label{Choosing_alpha}

Another implementation decision regards which significance level to choose. While in most cases assessments would lead to similar conclusions for any $\alpha$, there are some cases in which the choice of $\alpha$ may lead to different conclusions. 

To  illustrate that, we consider the empirical application from \cite{Dix}, which is analyzed in Section \ref{Sec: SS illustrations}. However, we focus now on the weighted OLS specification. We consider the design-based assessment for this specification to evaluate whether the inference method proposed by \cite{Adao}, with the null imposed, is reliable. When we impose the null, instead of using the residuals from the shift-share regression to estimate the standard errors, we use the residuals from a regression that imposes the null hypothesis to calculate the residuals (see \cite{Adao} for details). 

\begin{figure}[H] 

\begin{center}
\caption{{\bf Assessment of \cite{Adao} inference method with the null imposed}} \label{Fig_SS}

\begin{tabular}{c}

\includegraphics[scale=1]{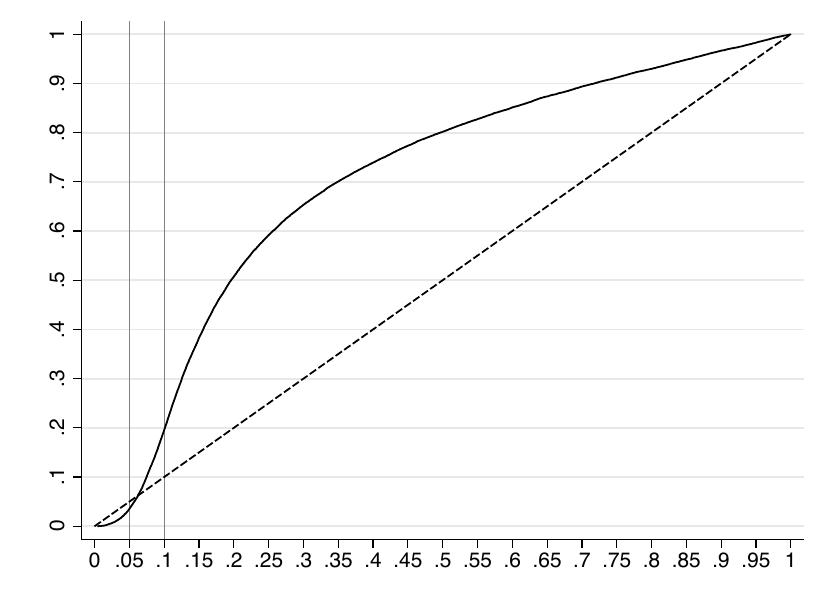}

\end{tabular}

\end{center}

\small{Notes: This figure presents the CDF of the p-values in the simulations used to construct the design-based assessment for the inference method proposed by \cite{Adao} with the null imposed. We consider the weighted OLS specification from \cite{Dix} (which is also analyzed in column 6 of Appendix Table \ref{Table_SS_CRVE}). The dashed line is the CDF of an uniform $[0,1]$ random variable. }

\end{figure}

When we consider an assessment with $\alpha = 5\%$, the design-based assessment is smaller than $5\%$ (at 3.5\%) which, at a first glance, could indicate that this inference method is actually conservative. However, when we consider $\alpha = 10\%$, then the design-based assessment becomes substantially bigger than $\alpha$ (at 20\%), suggesting relevant over-rejection. In Figure \ref{Fig_SS}, we show the CDF of the p-values in the simulations used in the assessments. If the inference method were working well, then we should expect it to be approximately the CDF of a uniform variable $[0,1]$. In contrast, we see  under-rejection for lower values of $\alpha$ and over-rejection for higher values. 

The source of this problem is that we are using an inference method in which we estimate the standard errors imposing the null hypothesis. Consider a very simple example to illustrate this issue. We have an iid sample of $N$ observations from $Z_i \sim N(\mu,\sigma^2)$, and we want to test the null $\mu = 0$. A t-statistic with standard errors calculated without imposing the null is $t =  \hat \mu /\widehat{se}$, where $\hat  \mu = \frac{1}{N}\sum_{i=1}^N Z_i$, $\widehat{se} = \sqrt{\hat \sigma^2/N}$ and $\hat \sigma^2={\frac{1}{N-1}\sum_{i=1}^N(Z_i - \hat \mu)^2}$. It is well-known that, in this simple example, $\hat \mu$ and $\widehat{se}$ are independent under normality. In contrast, when we calculate the standard errors imposing the null, we use $\widehat{se}_{\mbox{\tiny null}} = \sqrt{\hat \sigma_{\mbox{\tiny null}}^2/N}$, where $\hat \sigma_{\mbox{\tiny null}}^2={\frac{1}{N}\sum_{i=1}^N(Z_i - 0)^2}$. Therefore,  $\widehat{se}_{\mbox{\tiny null}}^2 = \frac{N-1}{N}\widehat{se}^2 + \frac{\hat \mu^2}{N}$, showing that, for a given $N$, there will be a positive correlation between $\widehat{se}_{\mbox{\tiny null}}$ and $|\hat \mu|$. This creates a downward bias in the rejection rates, that is stronger when $\alpha$ is smaller, because we generally need larger values of $|\hat \mu|$ to reject the null at lower levels of significance $\alpha$. As a consequence, we may have situations in which an assessment is close to $\alpha$ (or even smaller) when $\alpha = 5\%$, but it indicates large over-rejection when $\alpha = 10\%$.

Therefore, we recommend that applied researchers construct assessments using different significance levels (for example, 1\%, 5\%, and 10\%), particularly when assessing inference methods that impose the null. This novel insight is important not only for inference assessments, but also for MC simulations and EMCS in general.

\section{Concluding Remarks/Recommendations}
\label{Recommendations}

We analyze the use of different types of simulations that applied researchers may use to assess their inference methods. We discuss trade-offs between different alternatives that have not been previously considered in the literature, and present novel results related to the implementation of simulations to assess inference methods.

In light of our results, we recommend that, whenever it is reasonable to consider that outcomes are stochastic, applied researchers should \emph{at the very least} assess their inference methods using a simple assessment, such as the one replacing outcomes with iid normals, $\Gamma(\Phi(.))$.\footnote{As discussed in Section \ref{Source_uncertainty}, model-based assessments holding covariates fixed can be informative about both conditional and unconditional inference. } This can be seen as a compromise between standard MC simulations and more complex assessments such as EMCS, in that it holds the covariates matrix as in the data, but does not aim to estimate the distribution of the errors using the data. This assessment is straightforward, virtually costless in terms of implementation, and provides a valuable baseline check that is immune to sequential-testing concerns or cherry-picking. While it cannot capture all possible issues, we show evidence that it can still reveal important problems  and serve as a minimal diagnostic tool that nearly every applied researcher can and should employ. Importantly, applied researchers should have a clear understanding about potential problems that $\Gamma(\Phi(.))$ would not be able to detect, particularly in settings in which $\Gamma(\Phi(.))$  is close to $\alpha$.

At the same time, more complex assessments may also be useful in some applications, as they may approximate richer features of the empirical setting and potentially detect problems that $\Gamma(\Phi(.))$ would miss. However, they may come with higher implementation costs, may introduce  complications  such as sequential-testing distortions, and may lead to misleading conclusions depending on how they are implemented. We also show evidence that more complex assessments may have bad finite-sample properties which may prevent them from identifying problems that even $\Gamma(\Phi(.))$ would detect,  as illustrated in Section \ref{Sec_DID_N1=1}. Therefore, given its low implementation cost, we see value in considering $\Gamma(\Phi(.))$ as an additional check even when a more complex assessment is being considered.

Overall, we view the decision of whether to adopt more complex assessments---and, in particular, which assessment to adopt---as application-specific. Differently from $\Gamma(\Phi(.))$, more complex assessments may not be straightforward to choose and implement, and they may introduce additional complications, such as sequential-testing problems. Therefore, researchers need to carefully analyze the properties of the assessments in their particular setting and weight their respective advantages and disadvantages. For instance,  we show in Section 3 that the implementation of a design-based assessment commonly considered in shift-share designs may lead to misleading results, while modified versions of it can provide useful information about the validity of different inference methods in this setting. In this setting,  $\Gamma(\Phi(.))$ would be less informative. More broadly, evaluating the suitability of different types of complex assessments in other empirical contexts remains an interesting direction for future research.

We hope to encourage applied researchers to more routinely use simulations as sanity checks in their empirical applications, with a better understanding on how to choose, implement and interpret these simulations. This has the potential  of substantially improving the reliability of scientific evidence, even if we consider assessments that are straightforward and  can be implemented at a very low cost by applied researchers.



\singlespace

\renewcommand{\refname}{References} 

\bibliographystyle{apalike}
\bibliography{bib.bib}

\pagebreak

\pagebreak

\begin{table}[H]

 \centering
\caption{{\bf Assessments for DiD with $N_1=1$}} \label{Table_DiD_N1=1}
 \begin{lrbox}{\tablebox}
 \begin{tabular}{lccccc}

\cline{1-6}
\cline{1-6}

& median & p10 & p90 & $\mbox{Pr}(\Gamma>0.1)$ & $\mathbb{E}[\widehat \beta | \Gamma < 0.1]$ \\

 & (1) & (2) & (3) & (4) & (5) \\ 
 
\cline{1-6}

\multicolumn{6}{c}{Panel A: Assessments for cluster-robust standard errors} \\

$\Gamma_{\mbox{\tiny cluster}}(\Phi(.))$ & 0.787 & 0.787 & 0.787 & 1.000 & - \\ 
\\
$\Gamma_{\mbox{\tiny cluster}}(\widetilde F_{\mbox{\tiny rb}}(.;  \mathbf{Y}))$ & 0.774 & 0.717 & 0.827 & 1.000 & - \\ 
\\
$\Gamma_{\mbox{\tiny cluster}}(\widetilde F_{\mbox{\tiny npb}}(.;  \mathbf{Y}))$ & 0.054 & 0.045 & 0.063 & 0.000 & -0.004 \\ 
 &  &  &  &  & [0.009] \\

 \\

 \multicolumn{6}{c}{Panel B: Assessments for robust standard errors} \\

$\Gamma_{\mbox{\tiny robust}}(\Phi(.))$ & 0.075 & 0.075 & 0.075 & 0.000 & 0.000 \\
 \\
 \\
$\Gamma_{\mbox{\tiny robust}}(\widetilde F_{\mbox{\tiny rb}}(.;  \mathbf{Y}))$ & 0.168 & 0.108 & 0.238 & 0.930 & 0.027 \\
 &  &  &  &  & [0.032] \\
 \\
$\Gamma_{\mbox{\tiny robust}}(\widetilde F_{\mbox{\tiny npb}}(.;  \mathbf{Y}))$ & 0.000 & 0.000 & 0.000 & 0.000 & -0.004 \\
 &  &  &  &  & [0.009] \\

\cline{1-6}
\cline{1-6}
\end{tabular}
 \end{lrbox}
\usebox{\tablebox}\\
\settowidth{\tableboxwidth}{\usebox{\tablebox}} \parbox{\tableboxwidth}{\footnotesize{ Notes: this table presents the results for different types of assessments considered in Section \ref{Sec_DID_N1=1}. All assessments are constructed based on tests with nominal level of 5\%. The assessment based on iid normals does not depend on the draw of the original data, so we report the rejection rate from 10,000 simulations using iid normals as the outcome variable. For the other assessments, we generate $5000$ draws of the original data, and then for each draw we construct the assessment using $1000$ simulations. Columns 1 to 3 present the median, 10th percentile and 90th percentile of the assessments (across realizations of the original data). Column 4 presents the probability that the assessment is greater than 0.1, while column 5 presents the bias of the estimator conditional on an assessment smaller than 0.1. The choice of 0.1 as the threshold for flagging that the inference assessment is unreliable is arbitrary. Other choices lead to qualitatively similar results. }
}

\end{table}

\begin{table}[H]

 \centering
\caption{{\bf Assessments for DiD with $W_i|D_i=1 \sim $ log-normal (true size $=14\%$) }} \label{Table_lognormal}
 \begin{lrbox}{\tablebox}
 \begin{tabular}{lccccc}

\cline{1-6}
\cline{1-6}

& median & p10 & p90 & $\mbox{Pr}(\Gamma>0.1)$ & $\mathbb{E}[\widehat \beta | \Gamma < 0.1]$ \\

 & (1) & (2) & (3) & (4) & (5) \\ 
 
\cline{1-6}

$\Gamma_{\mbox{\tiny cluster}}(\Phi(.))$ & 0.070 & 0.070 & 0.070 & 0.000 & 0.000 \\
 \\
 
$\Gamma_{\mbox{\tiny cluster}}(\widetilde{F}_{\mbox{\tiny rb}}(.;\mathbf{Y}))$ & 0.066 & 0.062 & 0.071 & 0.000 & 0.010 \\
 & & & & & [0.007] \\
 
$\Gamma_{\mbox{\tiny cluster}}(\widetilde{F}_{\mbox{\tiny npb}}(.;\mathbf{Y}))$ & 0.099 & 0.072 & 0.202 & 0.491 & -0.145 \\
 & & & & & [0.008] \\
 
$\Gamma_{\mbox{\tiny cluster}}(\widetilde{F}_{\mbox{\tiny WGAN1}}(.;\mathbf{Y})))$  & 0.090 & 0.071 & 0.141 & 0.355 & -0.122 \\
 & & & & & [0.007] \\
 
$\Gamma_{\mbox{\tiny cluster}}(\widetilde{F}_{\mbox{\tiny WGAN2}}(.;\mathbf{Y})))$ & 0.069 & 0.063 & 0.076 & 0.000 & -0.007 \\
 & & & & & [0.007] \\

\cline{1-6}
\cline{1-6}

\end{tabular}
 \end{lrbox}
\usebox{\tablebox}\\
\settowidth{\tableboxwidth}{\usebox{\tablebox}} \parbox{\tableboxwidth}{\footnotesize{ Notes: this table presents the results for different types of assessments considered in Section \ref{Sec: DiD lognormal}. All assessments are constructed based on tests with nominal level of 5\%. The assessment based on iid normals does not depend on the draw of the original data, so we report the rejection rate from 10,000 simulations using iid normals as the outcome variable. For the other assessments, we generate $5000$ draws of the original data, and then for each draw we construct the assessment using $1000$ simulations. Columns 1 to 3 present the median, 10th percentile and 90th percentile of the assessments (across realizations of the original data). Column 4 presents the probability that the assessment is greater than 0.1, while column 5 presents the bias of the estimator conditional on an assessment smaller than 0.1. The choice of 0.1 as the threshold for flagging that the inference assessment is unreliable is  arbitrary. Other choices lead to qualitatively similar results. }
}

\end{table}

\begin{table}[H]

\begin{footnotesize}
 \centering
\caption{{\bf Design-based Assessments for shift-share designs}} \label{Table_SS}
 \begin{lrbox}{\tablebox}
\begin{tabular}{lcccccccc}
\cline{1-9}
\cline{1-9}

&\multicolumn{2}{c}{China shock} & & \multicolumn{2}{c}{Exposure to robots} & & \multicolumn{2}{c}{Trade liberalization} \\ \cline{2-3} \cline{5-6} \cline{8-9} 

& Main effects & Placebo && Main effects & Placebo & & Main effects & Placebo \\

 & (1) & (2) & & (3) & (4) & & (5) & (6) \\ 
 
\cline{1-9}

\multicolumn{9}{c}{Panel A: design-based assessment for cluster-robust standard errors} \\
 \\

Cluster-robust & 0.412 & 0.237 & & 0.330 & 0.059 & & 0.313 & 0.001 \\
 \\

\multicolumn{9}{c}{Panel B: $\boldsymbol{\epsilon}$-fixed design-based assessment for cluster-robust standard errors} \\
\\
Cluster-robust & 0.317 & 0.226 & & 0.256 & 0.049 & & 0.012 & 0.000 \\
 \\

\multicolumn{9}{c}{Panel C: design-based assessments for new inference methods} \\
\\
\cite{Adao} & 0.076 & 0.098 & & 0.355 & 0.221 & & 0.547 & 0.511 \\
 \\
\cite{Adao} & 0.042 & 0.041 & & 0.296 & 0.107 & & 0.220 & 0.116 \\

(null imposed) \\
\\
 
\# of clusters & 48 & 48 & & 48 & 48 & & 91 & 91 \\
 
\# of observations & 772 & 772 & & 722 & 722 & & 411 & 411 \\
 
\# of sectors & 395 & 395 & & 19 & 19 & & 20 & 20 \\

\cline{1-9}
\cline{1-9}

\end{tabular}
 \end{lrbox}
\usebox{\tablebox}\\
\settowidth{\tableboxwidth}{\usebox{\tablebox}} \parbox{\tableboxwidth}{\footnotesize{ Notes: this table presents different types of assessments for the shift-share applications analyzed in Section \ref{Sec: SS illustrations}. We consider both specifications with the main outcome of interest in the original papers (columns 1, 3, and 5), and in which the outcome variable is a placebo (columns 2, 4, and 6). Panels A and C consider a design-based assessment holding $\mathbf{y}$ constant. In Panel A we assess inference based on cluster-robust standard errors with HC3 correction, while in Panel C we assess the new inference methods proposed by \cite{Adao} (both the standard version and the version with the null imposed). In Panel B, we consider the $\boldsymbol{\epsilon}$-fixed design-based assessment discussed in Section \ref{Sec: DB in SS} to assess inference based on cluster-robust standard errors with HC3 correction. In all cases, assessments are based on random draws of iid standard normal shocks, where we calculate the rejection rate for a 5\%-level test. In column 1, we present results from the specifications considered by \cite{Adao} in their Section II, which is based on the application from \cite{Autor}. In column 2, we present the same results, but using a pre-treatment outcome (variation from 1980 to 1990). In columns 3 and 4 we present the assessments for a specification for the main effects and for a placebo specification from \cite{Acemoglu}. Finally, in columns 5 and 6 we present a specification for the main effects and a placebo specification from \cite{Acemoglu}. In all cases, we consider unweighted OLS regressions. }
}

\end{footnotesize}

\end{table}


\pagebreak

\pagebreak

\appendix

\setcounter{table}{0}
\renewcommand\thetable{A.\arabic{table}}

\setcounter{figure}{0}
\renewcommand\thefigure{A.\arabic{figure}}

\section{Online Appendix}

\onehalfspacing

\subsection{Design-based assessments - theory and simulations} 

\subsubsection{Proof of Proposition \ref{Prop}}
\label{Appendix_design_based}

Consider a shift-share designs setting with $i=1,...,N$ regions that are subject to $f=1,...,F$ aggregate shocks $\mathcal{X}_f$. The shift-share variable is given by ${x}_i = \sum_{f=1}^F w_{if}\mathcal{X}_f$, where the shares $w_{if}$ reflect how shock $\mathcal{X}_f$ affects unit $i$. The regression model is then given by
\begin{eqnarray}
y_i = \gamma + \beta x_i + \eta_i,
\end{eqnarray}
where $\beta$ is the effect of $x_i$ on $y_i$. Let $\mathbf{y}$ be the $N \times 1$ vector of outcomes.

Observations $i=1,...,N$ are partitioned into equally-sized groups $\Lambda_1,...,\Lambda_F$, with $w_{if} = 1$ if $i \in \Lambda_f$, and $w_{if} = 0$ otherwise. Assume also that $\mathcal{X}_f \in \{0,1\}$ and $ \sum_{f=1}^F \mathcal{X}_f =F/2$. We assume the vectors  $\{\epsilon_i\ : i \in \Lambda_f\}$ are iid across $f$.

We consider an inference assessment with random draws of $\widetilde{\mathcal{X}}_f \in \{0,1\}$ such that $\sum_{f=1}^F \widetilde{\mathcal{X}}_f =F/2$, while holding $\mathbf{y}$ fixed.  More specifically, for each draw of $\widetilde{\mathcal{X}}_f \in \{0,1\}$, we run the regression of $y_i = \gamma_0 + \gamma \tilde x_i + \tilde \eta_i$, where $\tilde x_i = \sum_{f=1}^F w_{if} \tilde{\mathcal{X}_f}$, yielding $\hat \gamma^b$. Then we test the null $\gamma=0$ using robust standard errors at a significance level $\alpha$, and the assessment is the rejection rate in these simulations. The DGP in the simulations set  potential outcomes as $\tilde y_i(0) = \tilde y_i(1) = y_i$ (which are fixed, given the sampling framework of the simulations) and the distribution for $\widetilde{\mathcal{X}}_f $ described above.\footnote{Given the structure of this shift-share design example, we only need to define the potential outcomes $y_i(x)$ for $x \in \{0,1\}$.} Therefore, the null hypothesis $H_0:\gamma=0$ is true in this DGP.

Uncertainty in these simulations comes only from realizations of $ \widetilde{\mathcal{X}}_f$. Let $\mathbb{E}^\ast[.| \mathbf{y}]$ and $\mathbb{V}^\ast(. | \mathbf{y})$ denote the expectation and variance operators with respect to  this measure, conditional on $\mathbf{y}$. From Lemma 5 from \cite{IK}, $\mathbb{E}^\ast [\hat \gamma^b | \mathbf{y} ] =0$, so  the estimator $\hat \gamma^b$ is unbiased. Let $\mathbb{V}^\ast_{\mbox{\tiny true}} \equiv \mathbb{V}^\ast\left(\hat \gamma^b | \mathbf{y} \right)$ be the true variance of $\hat \gamma^b$ in these design-based simulations. Note that this is a number for a fixed $\mathbf{y}$, and a random variable depending on the errors $\epsilon_i$ when $\mathbf{y}$ is treated as a random variable. Also, let $\mathbb{V}^\ast_{\mbox{\tiny robust}}$ be the true variance in case treatment were assigned at the individual in the design-based simulation. This is what the robust standard errors would asymptotically recover in these simulations when $N \rightarrow \infty$.

\begin{proposition} 
    Consider the shift-share design setting described in this section, and assume the vectors $\{\epsilon_i: i \in \Lambda_f\}$ are iid across $f$ with $\mathbb{E}[\epsilon_i]=0$, $\mathbb{V}(\epsilon_i)=\sigma^2$, and $cov(\epsilon_i,\epsilon_s)=\rho$ for $i \neq s$ and $i,s \in \Lambda_f$ for some $f$. Consider an asymptotic sequence in which $F \rightarrow \infty$ where we maintain that each group $\Lambda_f$ has exactly $m$ observations (so $N = m \times F$) and $\sum_{f=1}^F \mathcal{X}_f=F/2$. Then 
    \begin{eqnarray}
        \frac{\mathbb{V}^\ast_{\mbox{\tiny robust}}}{\mathbb{V}^\ast_{\mbox{\tiny true}}} \overset{a.s.}{\rightarrow} \frac{\beta^2 + 4 \sigma^2}{m\beta^2 + 4 \sigma^2 + 4(m-1)\rho}.
    \end{eqnarray}
\end{proposition}

\begin{proof}

Let $\bar\epsilon_f:=m^{-1}\sum_{i\in\Lambda_f}\epsilon_i$ and
$\bar\epsilon:=N^{-1}\sum_{i=1}^N\epsilon_i=F^{-1}\sum_{f=1}^F\bar\epsilon_f$.
Let $\mathcal{T}\subset\{1,\dots,F\}$ denote the set of treated groups in the original data ($\mathcal{X}_f=1$), with $|\mathcal{T}|=F/2$.
Write $s_f:=\mathbf{1}\{f\in\mathcal{T}\}-\tfrac12\in\{-\tfrac12,\tfrac12\}$ and
$t_i:=\mathbf{1}\{i\in\Lambda_f \text{ for some } f\in\mathcal{T}\}-\tfrac12\in\{-\tfrac12,\tfrac12\}$.
Note that $\sum_{f=1}^F s_f=0$, $\sum_{f=1}^F s_f^2=F/4$, $\sum_{i=1}^N t_i=0$, and $\sum_{i=1}^N t_i^2=N/4$.

By Lemma 5 and Lemma 2 of \cite{IK}, we can write
\begin{align}
\mathbb{V}^\ast_{\mbox{\tiny true}} 
&= \frac{4}{F(F-2)}\sum_{f=1}^F \Big(\beta s_f + (\bar\epsilon_f-\bar\epsilon)\Big)^2, \label{eq:Vtrue}\\
\mathbb{V}^\ast_{\mbox{\tiny robust}} 
&= \frac{4}{N(N-2)}\sum_{i=1}^N \Big(\beta t_i + (\epsilon_i-\bar\epsilon)\Big)^2. \label{eq:Vrob}
\end{align}

Now note that: \[
\frac{1}{F}\sum_{f=1}^F\!\Big(\beta s_f + (\bar\epsilon_f-\bar\epsilon)\Big)^2
= \beta^2\Big(\frac{1}{F}\sum_{f=1}^F s_f^2\Big)
+ \frac{2\beta}{F}\sum_{f=1}^F s_f(\bar\epsilon_f-\bar\epsilon)
+ \frac{1}{F}\sum_{f=1}^F(\bar\epsilon_f-\bar\epsilon)^2.
\]

Since $|\mathcal{T}|=F/2$, $\tfrac{1}{F}\sum s_f^2=\tfrac14$ exactly. For the cross term,
\[
\frac{1}{F}\sum_{f=1}^F s_f(\bar\epsilon_f-\bar\epsilon)
= \frac{1}{2F}\!\sum_{f\in\mathcal{T}}\!\bar\epsilon_f
  - \frac{1}{2F}\!\sum_{f\notin\mathcal{T}}\!\bar\epsilon_f
  - \bar\epsilon\cdot \frac{1}{F}\sum_{f=1}^F s_f
= \frac14\Big(\overline{\bar\epsilon}_{\,\mathcal{T}}-\overline{\bar\epsilon}_{\,\mathcal{T}^c}\Big),
\]
where $\overline{\bar\epsilon}_{\,\mathcal{T}}=(F/2)^{-1}\sum_{f\in\mathcal{T}}\bar\epsilon_f$ and similarly for $\mathcal{T}^c$.
Because $\{\bar\epsilon_f\}_f$ are i.i.d. with $\mathbb{E}[\bar\epsilon_f]=0$ and
$\mathbb{E}[|\bar\epsilon_f|]<\infty$, the strong law of large numbers (SLLN) gives
$\overline{\bar\epsilon}_{\,\mathcal{T}} \overset{a.s.}{\rightarrow} 0$ and $\overline{\bar\epsilon}_{\,\mathcal{T}^c} \overset{a.s.}{\rightarrow} 0$.
For the last term,
\[
\frac{1}{F}\sum_{f=1}^F(\bar\epsilon_f-\bar\epsilon)^2
= \frac{1}{F}\sum_{f=1}^F \bar\epsilon_f^{\,2} - \bar\epsilon^{\,2}
\;\overset{a.s.}{\rightarrow}\; \mathbb{E}[\bar\epsilon_f^{\,2}] - 0
= \mathbb{V}[\bar\epsilon_f]=\frac{\sigma^2 + (m-1)\rho}{m},
\]
using the SLLN for the i.i.d.\ sequence $\{\bar\epsilon_f^2\}_f$ and $\bar\epsilon \overset{a.s.}{\rightarrow} 0$.
Hence,
\begin{equation}
\frac{1}{F}\sum_{f=1}^F\Big(\beta s_f + (\bar\epsilon_f-\bar\epsilon)\Big)^2
\;\overset{a.s.}{\rightarrow}\; \frac{\beta^2}{4} + \frac{\sigma^2+(m-1)\rho}{m}.
\label{eq:limit-F}
\end{equation}

Analogously, for the unit-level expression,
\[
\frac{1}{N}\sum_{i=1}^N\Big(\beta t_i + (\epsilon_i-\bar\epsilon)\Big)^2
= \beta^2\Big(\frac{1}{N}\sum_{i=1}^N t_i^2\Big)
+ \frac{2\beta}{N}\sum_{i=1}^N t_i(\epsilon_i-\bar\epsilon)
+ \frac{1}{N}\sum_{i=1}^N(\epsilon_i-\bar\epsilon)^2.
\]
Again $\frac{1}{N}\sum t_i^2=\frac14$ exactly. The cross term equals
$\frac14(\bar\epsilon_{\mathrm{tr}}-\bar\epsilon_{\mathrm{ct}})$, where
$\bar\epsilon_{\mathrm{tr}}=(N/2)^{-1}\sum_{i:\,t_i=1/2}\epsilon_i$ and
$\bar\epsilon_{\mathrm{ct}}=(N/2)^{-1}\sum_{i:\,t_i=-1/2}\epsilon_i$; by the SLLN, both converge a.s.\ to $0$.
Finally,
$\frac{1}{N}\sum_{i=1}^N(\epsilon_i-\bar\epsilon)^2
= \frac{1}{N}\sum_{i=1}^N \epsilon_i^2 - \bar\epsilon^{\,2}\overset{a.s.}{\rightarrow} \sigma^2.$
Therefore,
\begin{equation}
\frac{1}{N}\sum_{i=1}^N\Big(\beta t_i + (\epsilon_i-\bar\epsilon)\Big)^2
\;\overset{a.s.}{\rightarrow}\; \frac{\beta^2}{4} + \sigma^2.
\label{eq:limit-N}
\end{equation}

Combining Equations \ref{eq:limit-F} and \ref{eq:limit-N},
\[
\frac{\mathbb{V}^\ast_{\mbox{\tiny robust}} }{\mathbb{V}^\ast_{\mbox{\tiny true}} }
= \frac{F-2}{N-2}\cdot
\frac{\frac{1}{N}\sum_{i=1}^N (\beta t_i + (\epsilon_i-\bar\epsilon))^2}
     {\frac{1}{F}\sum_{f=1}^F (\beta s_f + (\bar\epsilon_f-\bar\epsilon))^2}
\;\overset{a.s.}{\rightarrow}\;
\frac{1}{m}\cdot \frac{\frac{\beta^2}{4}+\sigma^2}{\frac{\beta^2}{4}+\frac{\sigma^2+(m-1)\rho}{m}}
= \frac{\beta^2+4\sigma^2}{m\beta^2+4\sigma^2 + 4(m-1)\rho}.
\] \end{proof}

\subsubsection{Simulations for design-based assessments}
\label{Appendix_permutation_shocks}

We present simulations that illustrate the use to design-based assessments for assessing whether inference is distorted due to spatial correlation. Consider a setting in which we observe $Y_{is}$ for individual $i$ in state $s$. We have $N$ states, where half of them receive a treatment $T_s$. Each state has $10$ individuals. We consider simulations in which $Y_{is}(0) = \omega \xi_s + \epsilon_{is}$ and $Y_{is}(1) = \beta + Y_{is}(0)$, where $\epsilon_{is}$ is iid $N(0,1)$ across $i$ and $s$, and $ \xi_s$ is iid $N(0,1)$ across $s$. The parameter $\beta$ reflects the true treatment effect, while $\omega$ reflects the relevance of state-level shocks. The goal is to construct design-based assessments to evaluate whether state-level shocks impose relevant distortions for inference based on robust standard errors. 

We consider a design-based assessment using permutations of $(T_1,\hdots,T_N)$, where for each permutation we estimate $\hat \beta^p$ and conduct inference using robust standard errors. The parameter of interest is the ATE for a superpopulation. As discussed in Section \ref{Source_uncertainty}, a design-based assessment can be used to assess the reliability of an unconditional test. We then discuss below the case in which the parameter of interest is the sample ATE. The simulation results are presented in Appendix Table \ref{Table_design_based}. We summarize the main conclusions.

\begin{enumerate}

\item As shown in Panels A, when there is a true treatment effect ($\beta \neq 0$), a design-based assessment would be larger than, for example, 0.1 (indicating relevant spatial correlation) with a high probability, even when there is no spatial correlation ($\omega=0$).

\item As also shown in Panel A, the $\boldsymbol{\epsilon}$-fixed design-based assessment would have a much lower probability of indicating relevant spatial correlation when $\omega=0$. When the number of states increases, the probability of incorrectly indicate spatial correlation problems goes to zero.

\item When there is spatial correlation ($\omega \neq 0$), both assessments would be larger than 0.1 with a high probability, correctly indicating that there are relevant spatial correlation problems. This probability is increasing in the number of states (Panels B and C). 

\item These assessments are more informative about spatial correlation problems than checking whether there are significant effects in placebo regressions using pre-treatment outcomes (Panel C). For example, in this setting with $\omega = 0.3$, a placebo regression would be significant at 5\% in 14\% of the time. In contrast, the design-based assessment would be greater than 0.1 around 74\% (91\%) of the time when $N=20$ ($N=100$). So we may have settings in which the p-value of a placebo regression would be large (which would not raise a red flag for the applied researcher), but these assessments would correctly indicate that there is a problem with a high probability.

\item These assessments would also be able to detect problems in case we have heterogeneous treatment effects at the state level (Panel E). {In this panel, we consider simulations in which $Y_{is}(0) = \epsilon_{is}$ (so there is no spatial correlation in the error for the potential outcomes when untreated), but $Y_{is}(1) = 0.4 \xi_s + Y_{is}(0)$, with $\xi_s \sim N(0,1)$. Therefore, the superpopulation ATE is still zero, but we have heterogeneous treatment effects for different states. }

\end{enumerate}

Now consider instead that the parameter of interest is the SATE (so inference is conditional on the potential outcomes). In this case, the test size would depend on the realization of the potential outcomes. Therefore, for some realizations of the potential outcomes we would have relevant size distortions, and for some we would not. The design-based assessment would correctly recover the exact size of the test (conditional on the potential outcomes) for the simulations in which $\beta = 0$, since $Y_{is}(1) = Y_{is}(0)$ in these simulations. However, when $\beta \neq 0$, the design-based assessment would tend to over-state the relevance of spatial correlation, in the same way as presented in Appendix Table \ref{Table_design_based}. Therefore, the main conclusion that a true treatment effect would over-state problems due to spatial correlation remains relevant in this case.

\subsection{Shift-share designs: probability of flagging spatial correlation} \label{Appendix_SS_applications}

In Section \ref{Sec: SS illustrations}, we discuss the use of design-based assessments to detect spatial correlation in shift-share designs. Here we consider the probability that those assessments would correctly flag spatial correlation problems in case they are actually present. 

We consider the following exercise. Let $w_{if} \geq 0$ be the shares of a given application. We construct an outcome vector given by $Y_i^\ast = Z_i + \gamma \sum_{f=1}^F w_{if}{\mathcal{X}}^\ast_f$, where $Z_i \sim N(0,1)$, while ${\mathcal{X}}^\ast_f \sim N(0,1)$ are random shocks that will \underline{not} be the shocks considered by the applied researcher to construct the shift-share variable. We can think of ${\mathcal{X}}^\ast_f$ as unobserved variables in the error term that might be spatially correlated. Since those unobservables have the same structure of shares as the real shocks that the applied researcher observes, this spatially correlated shocks may generate relevant over-rejection for inference based on cluster-robust standard errors cluster-robust standard errors cluster-robust standard errors, as discussed by \cite{Adao}. Note that spatial correlation will be stronger if $|\gamma|$ is larger.

We consider then that the applied researcher runs a shift-share regression using the shift-share variable $X_i = \sum_{f=1}^F w_{if}{\mathcal{X}}_f$ (where ${\mathcal{X}}_f \sim N(0,1)$ are the shocks that she observes), and $Y_i^\ast$ as the outcome variable. Again, all of those $N(0,1)$ variables are iid. Since $X_i $ and $Y_i^\ast$ are independent, the expected value of the shift-share estimator in this DGP is zero. 

We consider simulations of this model with the structure of the three applications considered in Section \ref{Sec: SS illustrations}. For each realization of these random variables (that generates $Y_i^\ast$ and $X_i$), we (i) estimate the shift-share regression, and test the null using cluster-robust standard errors; (ii) calculate the design-based assessment and the $\boldsymbol{\epsilon}$-fixed design-based assessment. We do that for a number of different values for $\gamma$.

For each $\gamma$, we first calculate the test size. For all empirical applications, this is close to 5\% when $\gamma=0$ (because there is no spatial correlation in this case). However, when $\gamma$ increases, then we start to have over-rejection. Then we calculate the proportion of times in which each of the assessments would flag that there is a spatial correlation problem. We define that as the assessment being greater than 0.1 (alternative thresholds leads to similar conclusions). We plot in Appendix Figure \ref{Fig_SS_flagging} the probability of flagging a problem as a function of the test size (which, in turn, is a function of $\gamma$).

For the application from \cite{Autor}, these assessments would have a large probability of detecting problems, when there is spatial correlation. For example, if the spatial correlation is such that the test size using cluster-robust standard errors is 17\%, there would be a 90\% probability of having assessments greater than 0.1. For the application from \cite{Acemoglu}, the probability of detecting problems is a bit lower. For example, if the spatial correlation is such that the test size is around 15\%, there would be a 60\% probability of having assessments greater than 0.1. If the spatial correlation is stronger, then the probability of detecting problems would be larger. Finally, we note that the probability of detecting problems is much smaller for the application from \cite{Dix}. When the test size is around 15\%, we would only have a 35\% probability of having a design-based assessment greater than 0.1.

In case the design-based assessments do not detect problems due to spatial correlation, we recommend that applied researchers consider this kind of simulations. This way, they can evaluate whether the assessments would have a high probability of detecting problems when there are meaningful spatial correlation problems.

\subsection{Appendix Figures and Tables}

\begin{figure}[H] 

\begin{center}
\caption{{\bf Probability of detecting spatial correlation problems}} \label{Fig_SS_flagging}

\begin{tabular}{ccc}

A: China shocks & B: Exposure to robots & C: Trade liberalization \\

\includegraphics[scale=0.37]{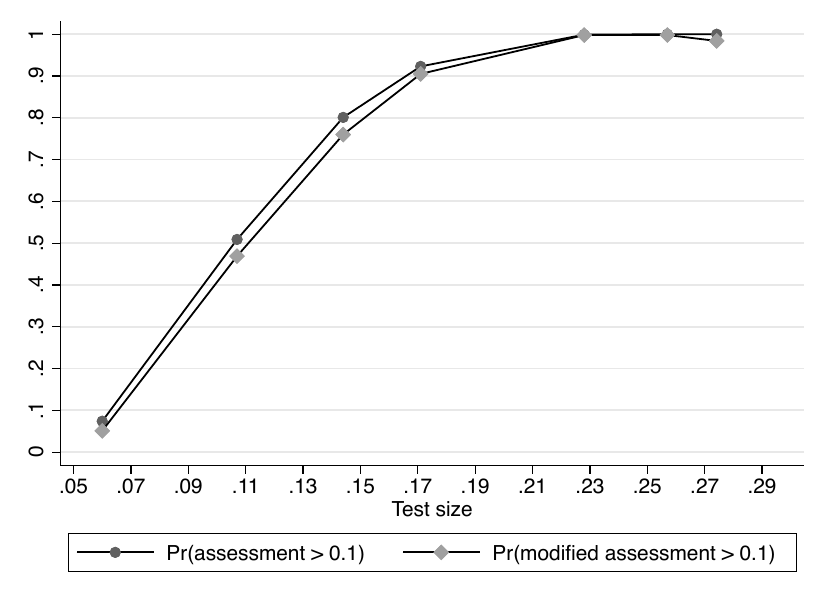} & \includegraphics[scale=0.37]{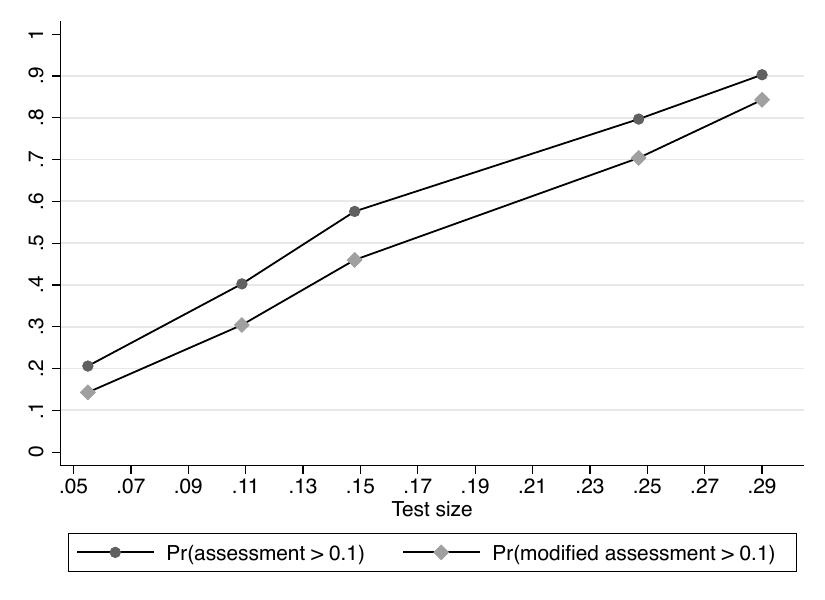} & \includegraphics[scale=0.37]{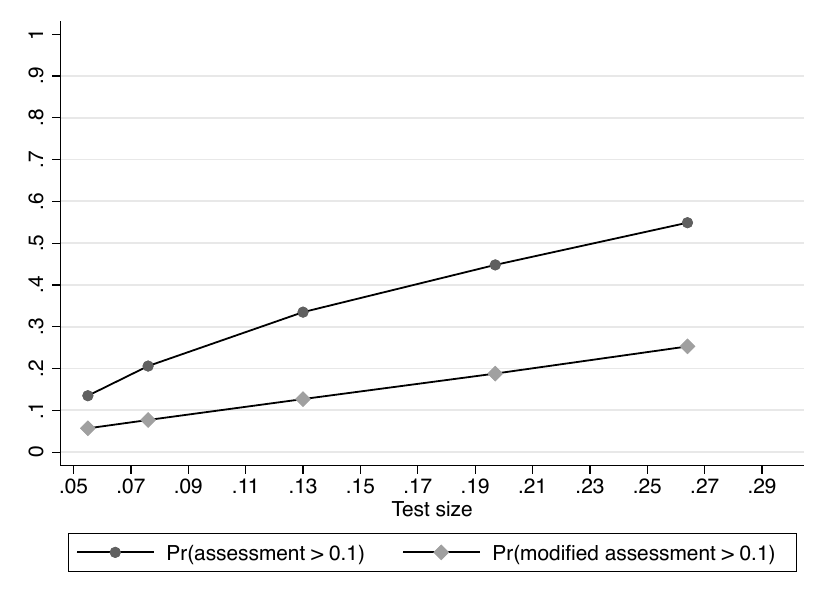}

\end{tabular}

\end{center}

\small{Notes: these figures present the results from the simulations described in Appendix \ref{Appendix_SS_applications}. Each point in the graphs represent one choice of $\gamma$, which determines the strength of the spatial correlation in the DGP used in the simulations. For that $\gamma$, we plot the implied rejection rate for inference using cluster-robust standard errors (which is increasing in $|\gamma|$), and the probabilities that the assessments are greater than 0.1. We consider the application from \cite{Autor} in Figure A, the one from \cite{Acemoglu} in Figure B, and from \cite{Dix} in Figure C. }

\end{figure}

\pagebreak

\begin{table}[H]

\begin{footnotesize}
 \centering
\caption{{\bf Simulations - design-based assessments }} \label{Table_design_based}
 \begin{lrbox}{\tablebox}
 \begin{tabular}{l C{3cm} C{3cm} C{3cm} }

\cline{1-4}

& Test size & $Pr(\Gamma_{\mathbf{y}} > 0.1)$  & $Pr(\Gamma_{\boldsymbol{\epsilon}} > 0.1)$  \\

 & (1)  & (2)  &  (3) \\ \cline{1-4}

\multicolumn{4}{c}{Panel A: positive treatment effect ($\beta=0.5$), but no spatial correlation ($\omega=0$)} \\
$N=20$ & 0.051 & 0.632 & 0.091 \\
$N=100$ & 0.049 & 0.715 & 0.008 \\
\\
\multicolumn{4}{c}{Panel B: positive treatment effect ($\beta=0.5$), and spatial correlation ($\omega=0.3$) } \\

$N=20$ & 0.140 & 0.927 & 0.688 \\
 
$N=100$ & 0.138 & 0.998 & 0.902 \\
\\
 
 \multicolumn{4}{c}{Panel C: no treatment effect ($\beta=0$), and spatial correlation ($\omega=0.3$) } \\
$N=20$ & 0.140 & 0.743 & 0.689 \\
 
$N=100$ & 0.138 & 0.913 & 0.902 \\
\\

 \multicolumn{4}{c}{Panel D: no treatment effect ($\beta=0$), and no spatial correlation ($\omega=0$) } \\
$N=20$ & 0.051 & 0.114 & 0.091 \\
$N=100$ & 0.049 & 0.009 & 0.008 \\
\\

 \multicolumn{4}{c}{Panel E: heterogeneous treatment effects } \\
$N=20$ & 0.129 & 0.672 & 0.615 \\

$N=100$ & 0.130 & 0.840 & 0.826 \\

\cline{1-4}

\end{tabular}
 \end{lrbox}
\usebox{\tablebox}\\
\settowidth{\tableboxwidth}{\usebox{\tablebox}} \parbox{\tableboxwidth}{\footnotesize{Notes: this table presents the results from the simulations discussed in Appendix \ref{Appendix_permutation_shocks}. Column 1 presents test size for inference based on robust standard errors, with a nominal level of 5\%. Column 2 (column 3) presents the probability that the $\mathbf{y}$-fixed design-based assessment ($\boldsymbol{\epsilon}$-fixed design-based assessment) is greater than 0.1. This would flag that inference based on robust standard errors might be problematic due to spatial correlation (all results remain similar if we consider alternative thresholds).  We run 20.000 simulations (draws of the original data) for each scenario, and for each draw the assessment is constructed using 500 permutations. }
}

\end{footnotesize}

\end{table}

\begin{table}[H]

\begin{footnotesize}
 \centering
\caption{{\bf Shift-share designs: assessments for alternative CRVE's }} \label{Table_SS_CRVE}
 \begin{lrbox}{\tablebox}
 \begin{tabular}{lcccccccc}
\cline{1-9} 

\cline{1-9}

&\multicolumn{2}{c}{China shock} & & \multicolumn{2}{c}{Exposure to robots} & & \multicolumn{2}{c}{Trade liberalization} \\ \cline{2-3} \cline{5-6} \cline{8-9} 

& unweighted & weighted && unweighted & weighted & & unweighted & weighted \\

 & (1) & (2) & & (3) & (4) & & (5) & (6) \\ 
 
\cline{1-9}

\multicolumn{9}{c}{Panel A: Model-based assessment (iid normal errors) } \\
 \\
CRVE & 0.073 & 0.093 & & 0.090 & 0.311 & & 0.062 & 0.146 \\
 \\
CRVE - HC3 & 0.059 & 0.019 & & 0.049 & 0.000 & & 0.056 & 0.031 \\
 \\

\multicolumn{9}{c}{Panel B: Model-based assessment (normal errors with estimated heteroskedasticity)} \\
\\
CRVE & 0.073 & 0.093 & & 0.087 & 0.310 & & 0.061 & 0.146 \\
 \\
CRVE - HC3 & 0.059 & 0.019 & & 0.047 & 0.000 & & 0.056 & 0.031 \\
 \\

\multicolumn{9}{c}{Panel C: Assessment with stochastic errors and shocks} \\
 \\
CRVE & 0.070 & 0.063 & & 0.067 & 0.123 & & 0.064 & 0.140 \\
 \\
CRVE - HC3 & 0.060 & 0.007 & & 0.042 & 0.000 & & 0.055 & 0.028 \\

\\
 
\# of clusters & 48 & 48 & & 48 & 48 & & 91 & 91 \\
 
\# of observations & 772 & 772 & & 722 & 722 & & 411 & 411 \\

\cline{1-9}
\end{tabular}
 \end{lrbox}
\usebox{\tablebox}\\
\settowidth{\tableboxwidth}{\usebox{\tablebox}} \parbox{\tableboxwidth}{\footnotesize{Notes: this table presents different types of assessments to assess the reliability of cluster-robust standard errors (CRVE) and cluster-robust standard errors with the HC3 correction (CRVE - HV3) for the empirical applications presented in Section \ref{Sec: SS illustrations}. Panel A and B considers model-based assessments. Panel A uses a distribution $\widetilde F(.;\mathbf{Y})$ iid normal, while Panel B estimates the variance of each observations as a function of population sizes, and considers a DGP with errors independent normal with this estimated heteroskedasticity. More specifically, we consider that $\epsilon_i \sim N(0,A+B/W_i) $, where $W_i$ represents the population of region $i$, and estimate the parameters $A$ and $B$ using the OLS residuals. Panel C considers an assessment in which the outcome is iid normal and shocks are iid normal (so the shift-share variable is also stochastic). Columns 1 and 2 refer to the application from \cite{Autor}. We consider the specification that \cite{Adao} used in their Table I. Column 1 uses unweighted OLS, while column 2 uses weighted OLS. Columns 3 and 4 refer to the application from \cite{Acemoglu}. Column 3 uses the specification from column 6 in their Table 2 (unweighted), while column 4 uses the specification from column 4 in their Table 2 (weighted). Columns 5 and 6 refer to the application from \cite{Dix}. Column 5 uses the specification from column 1 in their Table 2 (unweighted), while column 6 uses the specification from column 2 in their Table 2 (weighted). Overall, these results suggest that, in a setting with no spatial correlation, asymptotic approximations for the use of cluster-robust standard errors are less reliable for the weighted specifications. This is particularly true for the applications from \cite{Acemoglu} and \cite{Dix}. }
}

\end{footnotesize}

\end{table}

\pagebreak

\begin{table}[H]

\begin{footnotesize}
 \centering
\caption{{\bf Shift-share designs: point estimates and standard errors }} \label{Table_appendix}
 \begin{lrbox}{\tablebox}
 \begin{tabular}{lcccccccc}
\cline{1-9}
\cline{1-9}

&\multicolumn{2}{c}{China shock} & & \multicolumn{2}{c}{Exposure to robots} & & \multicolumn{2}{c}{Trade liberalization} \\ \cline{2-3} \cline{5-6} \cline{8-9} 

& Main effects & Placebo && Main effects & Placebo & & Main effects & Placebo \\

 & (1) & (2) & & (3) & (4) & & (5) & (6) \\ 
 
\cline{1-9}

Estimate & -0.504 & -0.110 & & -0.516 & -0.217 & & -1.976 & 0.727 \\
 \\
CRVE \\
~~~ Standard error & 0.103 & 0.048 & & 0.118 & 0.151 & & 0.822 & 1.096 \\
~~~ p-value & 0.000 & 0.023 & & 0.000 & 0.152 & & 0.016 & 0.508 \\
 \\
CRVE - HC3 \\
~~~ Standard error & 0.116 & 0.052 & & 0.150 & 0.226 & & 0.839 & 1.147 \\
~~~ p-value & 0.000 & 0.034 & & 0.001 & 0.338 & & 0.018 & 0.526 \\
 \\
\cite{Adao} \\
~~~ Standard error & 0.138 & 0.036 & & 0.053 & 0.070 & & 0.311 & 0.227 \\
~~~ p-value & 0.000 & 0.003 & & 0.000 & 0.002 & & 0.000 & 0.001 \\
 \\
\multicolumn{2}{l}{\cite{Adao} (null imposed)} \\
~~~ Standard error & 0.208 & 0.048 & & 0.115 & 0.342 & & 0.545 & 0.442 \\
~~~ p-value & 0.016 & 0.022 & & 0.000 & 0.525 & & 0.000 & 0.100 \\

\\

\# of clusters & 48 & 48 & & 48 & 48 & & 91 & 91 \\
 
\# of observations & 772 & 772 & & 722 & 722 & & 411 & 411 \\
 
\# of sectors & 395 & 395 & & 19 & 19 & & 20 & 20 \\

\cline{1-9}
\cline{1-9}

\end{tabular}
 \end{lrbox}
\usebox{\tablebox}\\
\settowidth{\tableboxwidth}{\usebox{\tablebox}} \parbox{\tableboxwidth}{\footnotesize{ Notes: this table presents the estimates, standard errors, and p-values when we consider inference based on cluster-robust standard errors (CRVE), cluster-robust standard errors with  HC3 correction (CRVE - HC3), and the procedure proposed by \cite{Adao} (without and with the null imposed). Columns 1 and 2 refer to the application from \cite{Autor}. We consider the specification that \cite{Adao} used in their Table I. Column 1 uses estimates for the main effects, while column 2 uses a placebo specification (considering variations in the outcome variable from 1980 to 1990). Columns 3 and 4 refer to the application from \cite{Acemoglu}. Column 3 uses the specification from column 6 in their Table 2, while column 4 uses the specification from column 4 in their Table 4. Columns 5 and 6 refer to the application from \cite{Dix}. Column 5 uses the specification from column 1 in their Table 2, while column 6 uses the specification from column 1 in their Table 4. }
}

\end{footnotesize}

\end{table}

\end{document}